\documentclass[%
 reprint,
 amsmath,amssymb,
 aps,
]{revtex4-1}

\usepackage{graphicx}% Include figure files
\usepackage{dcolumn}% Align table columns on decimal point
\usepackage{bm}% bold math
\usepackage{amssymb}
\usepackage{balance} 
\usepackage{bbold}
\usepackage{color}

\begin{document}

\preprint{APS/123-QED}

\title{Dynamics of suspensions of hydrodynamically structured particles: \\Analytic theory and experiment}% Force line breaks with \\

\author{Jonas Riest}
\email{j.riest@fz-juelich.de}
\affiliation{Forschungszentrum J\"ulich GmbH, ICS-3 - Soft Condensed Matter,\\ 52428 J\"ulich, Germany}

\author{Thomas Eckert}
\author{Walter Richtering}
\affiliation{
 Institute of Physical Chemistry, RWTH Aachen University, \\Landoltweg 2, 52056, Aachen, Germany
}%

\author{Gerhard N\"agele}
\affiliation{Forschungszentrum J\"ulich GmbH, ICS-3 - Soft Condensed Matter, \\52428 J\"ulich, Germany}

\date{\today}% It is always \today, today,
             %  but any date may be explicitly specified

\begin{abstract}
We present an easy-to-use analytic toolbox for the calculation of short-time transport properties of concentrated suspensions of spherical colloidal particles with internal hydrodynamic structure, and direct interactions described by a hard-core or soft Hertz pair potential. The considered dynamic properties include self-diffusion and sedimentation coefficients, the wavenumber-dependent diffusion function determined in dynamic scattering experiments, and the high-frequency shear viscosity. The toolbox is based on the hydrodynamic radius model (HRM) wherein 
the internal particle structure is mapped on a hydrodynamic radius parameter for unchanged direct interactions, and on an existing
simulation data base for solvent-permeable and spherical annulus particles. Useful scaling relations for the diffusion function and self-diffusion coefficient, known to be valid for hard-core interaction, are shown to apply also for soft pair potentials. We further discuss
extensions of the  toolbox to long-time transport properties including the low-shear zero-frequency viscosity and the long-time self-diffusion coefficient. 
The versatility of the toolbox is demonstrated by the analysis of a previous light scattering study of suspensions of non-ionic PNiPAM microgels
[Eckert \textit{et al.}, \textit{J. Chem. Phys.}, 2008, \textbf{129}, 124902] in which a detailed theoretical analysis of the dynamic data was left as an open task. 
By the comparison with Hertz potential based calculations, we show that the experimental data are consistently and accurately described using
the Verlet-Weis corrected Percus-Yevick structure factor as input, and for a solvent penetration length
equal to three percent of the excluded volume radius.
This small solvent permeability of the microgel particles has a significant dynamic effect at larger concentrations.
\end{abstract}

\maketitle

%\tableofcontents

\section{Introduction}
Suspensions of globular colloidal particles with internal hydrodynamic structure, and different surface boundary conditions, are abundant in soft matter science. Examples of technological and biomedical relevance
 are non-ionic and ionic microgel particles, and core-shell particles consisting of a dry spherical core and a shell of some soft material such as a polymer brush. These particulate systems are to a certain degree permeable to the solvent. Microgel particles
 in particular consist of a network formed by cross-linked polymer chains. They have useful features such as temperature-, pH-, salinity- and concentration-dependent \cite{Holmqvist2012} swelling behavior as well as elasticity and flexibility. This renders them as good
 candidates for various applications such as drug delivery agents \cite{Eichenbaum1999,Lopez2004,Sahiner2012}, the engineering of tissues \cite{Lally2007,Panda2008,Shen2012,Lyon2013}, and the modification of rheological properties \cite{Chari2013}. The elasticity
 of microgels can be controlled, e.g., by the amount of crosslinker, and the length of polymer chains used in the synthetization process. Microgels can be therefore considered as bridging the gap between genuine hard spheres and ultra-soft colloids \cite{Heyes2009}.
 
Although microgel and core-shell particle systems have been intensely studied experimentally over the past years, a quantitative theoretical description of the diffusion and rheological properties of concentrated suspensions is still on demand. This is owed
 to the complicated many-particle hydrodynamic interactions (HIs) which are significantly influenced by the hydrodynamic structure of the particles. A theoretical understanding of the influence of HIs on colloidal transport properties such as translational
 and rotational diffusion coefficients, the generalized sedimentation coefficient (hydrodynamic function), and high-frequency and zero-frequency viscosities is of key importance also in process engineering, e.g. in filtration and fractionation processes \cite{Roa,RichardBowen2001},
 and for the energy cost reduction in the transportation of colloidal suspensions through viscosity minimization. 
 
On a coarse-grained level, the porosity-averaged fluid flow inside a solvent-permeable particle is commonly described by the Brinkman-Debye-Bueche
 (BDB) equation invoking the Darcy permeability, $\kappa^2$, where $1/\kappa$ is the hydrodynamic penetration length \cite{Debye1948b,Brinkman1949}. Globular particles with an on average spherically symmetric hydrodynamic structure can be characterized by a
 permeability coefficient, $\kappa(d)$, depending on the radial distance, $d$, from the particle center \cite{Deutch1975}. Versatile hydrodynamic simulation tools such as the HYDROMULTIPOLE hydrodynamic force multipole method \cite{Cichocki1999} have been developed
 which allow for calculating transport properties of concentrated dispersions of hydrodynamically structured particles with the full inclusion of HIs. However, these simulations are numerically expensive, and in principle they must be performed separately for
 each particle model. In a recent series of papers, various short-time dynamic properties of dispersions of uniformly permeable spheres \cite{Abade2010c,Abade2010d,Abade2010,Abade2010a}, and of core-shell particles with uniformly permeable shell \cite{Abade2012,Cichocki2013,Cichocki2014},
 have been calculated using the HYDROMULTIPOLE method, as functions of particle concentration, reduced Darcy permeability, and shell-thickness to particle size ratio. The non-hydrodynamic direct particle interactions in these simulations have been taken for
 simplicity as pure hard-core interactions characterized by the excluded volume particle radius $a=\sigma/2$. Any softness in the effective pair potential, $V(r)$, between two globular particles at center-to-center distance $r$ is hereby disregarded. 
 
As discussed in \cite{Abade,Cichocki2013,Cichocki2014}, a simplifying concept allowing for abstracting from specific intraparticle structures is the so-called hydrodynamic radius model (HRM) which invokes the notion of an apparent no-slip hydrodynamic particle radius $a_h$ (see also 
 \cite{Anderson1991a,Anderson1996}). The HRM amounts to approximating a globular particle of spherically symmetric hydrodynamic structure by a no-slip sphere of hydrodynamic radius $a_h$ while leaving the effective pair potential unchanged. Under from an experimental
 viewpoint surprisingly general conditions, $a_h$ is unequivocally determined from the measurement of a single-particle transport property such as the translational diffusion coefficient $D_0^t$ or the intrinsic viscosity $[\eta]$. The definition of the HRM
 includes also spherical particles with fuzzy hydrodynamic structure and no sharp outer boundary, and with a soft pair potential such as for weakly cross-linked ionic microgels \cite{Holmqvist2012, Riest2012, Gottwald2004}. For spherical particles having excluded
volume interactions only with $a_h < a$, the HRM reduces to the so-called spherical annulus model. For the annulus model, numerically precise simulation results for various short-time dynamic properties have been given in \cite{Abade2012}. The good accuracy
 of the simplifying HRM was demonstrated in \cite{Abade2010,Cichocki2011,Abade2012} for uniformly permeable and core-shell spheres with pure excluded volume interactions, by a thorough comparison with simulation results. While a single hydrodynamic radius
 suffices to characterize the hydrodynamic intraparticle structure of many experimentally realized suspensions, regarding its influence on configuration-averaged transport properties, the replacement of the soft pair potential by an effective hard-core potential is in
 general a less successful strategy. Methods of calculating static suspension properties based on an effective hard-sphere potential such as the Barker-Henderson perturbation scheme, a second virial coefficient mapping, and additional variational methods commonly fail if the longer-ranged, soft part of the pair potential stretches out significantly beyond the physical excluded volume radius. For charge-stabilized colloids, e.g., this has been shown in \cite{McPhie2007,Nagele1996}. 
 
In this article, we present an easy-to-apply set of analytic methods, referred to for short as a toolbox, to calculate short-time transport properties of concentrated dispersions consisting of spherical colloidal particles with internal hydrodynamic structure, and with direct interactions 
given by the hard-core and two-parameter soft Hertz potentials. The 
latter potential is continuous and bounded, and it describes the energy penalty caused by the elastic deformation of two colliding spheres \cite{Riest2012,Likos2011a}. The Hertz potential
 has been shown to be a useful description of the direct interactions between non-ionic, soft microgels of low cross-link density, and this even though the particles are of a distinctly inhomogeneous structure \cite{Riest2012,Paloli2013}. Our toolbox is based
 on the HRM, and it takes advantage of the tabulated simulation data for spherical annulus particles listed in \cite{Abade2012}. The toolbox incorporates in particular useful approximate scaling relations for the
wavenumber-dependent sedimentation coefficient, $H(q)$, the short-time self-diffusion coefficient $D_S$, and the high-frequency viscosity $\eta_\infty$. These quantities are routinely determined in dynamic scattering experiments. The scaling relation expressions are known for permeable particles with hard-core interactions to be in remarkably good agreement with simulation
 data \cite{Abade2010c,Abade2010d,Abade2011}. We show that they apply likewise to particles with a soft pair potential, and we augment them by scaling expressions for the collective diffusion coefficient, $D_C$, and the associated sedimentation coefficient
 $K$. 
 
Moreover, we extend the toolbox to long-time dynamic properties of concentrated systems of hydrodynamically structured particles, including the low-shear zero-frequency suspension viscosity $\eta$, and the long-time translational self-diffusion coefficient $D_L$.
 Different from their short-time siblings, long-time transport properties are affected additionally by the non-instantaneous microstructural relaxation of the cloud of neighboring Brownian particles. This relaxation is controlled both by direct and hydrodynamic
 interactions. The toolbox extension to long-time properties combines the HRM with a factorization approximation method introduced originally by Medina-Noyola \cite{Medina-Noyola1988}, and  elaborated subsequently by Brady \cite{Brady1993,Brady1994} and Banchio {\em
 et al.} \cite{Banchio1999}. The HRM is useful also for long-time properties, since the hydrodynamic mobilities in the generalized many-particle Smoluchowski diffusion equation describing the configurational distribution function are time-independent
 \cite{Nagele1996}.
 
We demonstrate the accuracy of our user-friendly toolbox through the analysis of a light scattering study on a concentration series of non-ionic, submicron-sized PNiPAM microgel particles dispersed in dimethylformamide (DMF). We show that the
 static and dynamic scattering data for the static structure factor $S(q)$ and hydrodynamic function $H(q)$, respectively, can be quantitatively described in the complete experimental wavenumber range, on using a deduced solvent penetration length equal to three percent of the particle
 diameter. Calculations based on the Hertz potential show that the PNiPAM microgels behave statically as effective hard spheres in the whole fluid-phase concentration regime. 
 
The paper is organized as follows: The structure factor and the associated radial distributon function
 (RDF), $g(r)$, of the Hertz potential model are discussed in Subsec. \ref{subec:pair_correlations}, and compared with static light scattering data of non-ionic PNiPAM microgels. Subsec. \ref{subsec:hyd_structures} describes how hydrodynamic particle structures
 are related to the generic HRM, exemplified for the two important examples of uniformly permeable spheres and non-permeable rigid spheres with partial hydrodynamic surface slip. Sec. \ref{sec:short_time_dynamics} includes the discussion of various short-time dynamic
 properties, including $H(q)$, $D_S$, $K$, and the high-frequency viscosity $\eta_\infty$. Approximate analytic expressions for these quantities are presented, and the accuracy of these expressions is scrutinized against simulation results for the spherical annulus
 model. In Sec. \ref{sec:theory-experiment}, the toolbox results are compared with static and dynamic light scattering (SLS and DLS) data by Eckert {\em et al.} on PNiPAM microgel suspensions. Sec. \ref{sec:long-time-properties} describes the extension of the toolbox to long-time transport
 properties including the long-time translational self-diffusion coefficient $D_L$ and the zero-frequency 
viscosity $\eta$. Our conclusions are contained in Sec. \ref{sec:conclusions}. The pairwise additivity (PA) and the self-part corrected Beenakker Mazur methods of calculating
 short-time transport properties are explained in the Appendices A and B, respectively, in the context of the HRM.

\section{Equilibrium Microstructure and Hydrodynamic Radius Model}

\subsection{Static pair correlations}
\label{subec:pair_correlations}

Owing to the many possible applications, various theoretical schemes have been developed for the  analytic calculation of static properties of microgel suspensions. Progress in this direction was made, in particular for ionic microgels, through the development of effective pair potentials characterized by the suspension temperature and salinity, and the bare charge of the microgel particles \cite{Denton2000,Denton2003,Chung2013}. The validity of these effective pair potentials has been scrutinized in various joint theoretical-experimental studies (see e.g. \cite{Heinen2011,Riest2012}). In contrast to ionic microgels, the effective pair potentials used for non-ionic soft  
microgels where short-range interactions are not masked by the longer-ranged  
electrostatic repulsion are to date still on a more heuristic level. The pronounced dependence of the pair potential in non-ionic microgel systems on ambient and intra-particle conditions such as the solvent quality and temperature, number and distribution of crosslinker, and length and functionality of polymer chains requires  a larger number of parameters characterizing the effective pair potential on a microscopic level. On a more coarse-grained level, simplifying pair potentials have been used such as the hard-sphere \cite{Eckert2008} and elastic Hertz \cite{Riest2012,Palli2014} potentials, and certain ultra-soft pair potentials \cite{Heyes2009}. The intricate dependence  
on environmental parameters is hidden in these coarse-grained potentials in a reduced number of interaction parameters such as the effective interaction 
strength and the effective particle radius.

A useful coarse-grained effective pair potential for non-ionic globular microgel particles 
is the Hertz potential,  
\begin{equation}
\beta V(r) = \begin{cases} \epsilon\left(1-\frac{r}{\sigma_\text{s}}\right)^{\frac{5}{2}} & r\leq \sigma_\text{s} \\ 
0 & r> \sigma_\text{s}\,, \end{cases}
\label{eq:pot_hertz}
\end{equation}
which describes the elastic deformation of two spheres in contact. 
Here, $\beta=1/(k_{\text{B}}T)$ is the reduced inverse temperature with Boltzmann constant 
$k_{\text{B}}$ and absolute temperature $T$, and $\sigma_\text{s}$ plays the role of an effective soft particle diameter. For distances $r\geq \sigma_\text{s}$, two Hertz model particles do not interact with each other.  
The strength of the continuous potential is quantified by 
the non-dimensional elasticity parameter (effective potential strength), 
\begin{equation}
\epsilon = \frac{2Y\sigma_\text{s}^3}{15k_{\text{B}}T\left(1-\nu^2\right)} \,,
\label{eq:hertz_strength}
\end{equation}
depending on the bulk modulus $Y$ and the Poisson ratio $\nu$ 
of a particle \cite{Likos2011a,Riest2012}. On approximating the mesoscopic particle elastic moduli $Y$ and $\nu$ by macroscopic values, the estimate $\epsilon\approx10^4\sim10^5$ is obtained for micron-sized microgels. Note here the strong size dependence $\epsilon\propto\sigma_\text{s}^3$, implying a significantly decreased potential strength    
for smaller microgel particles. The Hertz potential has been shown to provide a nearly fit-parameter free description of the structure and phase behavior of neutral microgel systems, in  good agreement with experimental results for various particle sizes \cite{Riest2012, Paloli2013}, and values of $\epsilon$ in the range from $10^2 - 10^4$. 
While for large $\epsilon$ the Hertz and hard-sphere potentials lead to very similar  results for the equilibrium microstructure, the effective diameter in the Hertz potential is in general somewhat larger than the corresponding hard-sphere value, i.e.  
$\sigma_\text{s}\gtrsim\sigma$. This reflects the fact that the Hertz potential incorporates overall the softness of a microgel particle which in turn originates from the radially inhomogeneous crosslinker density.

For very large values of $\epsilon$, the Hertz potential is practically indistinguishable from the hard-sphere potential
\begin{equation}
V_\text{HS}(r) = \begin{cases} \infty & r\leq \sigma = 2 a\\ 
0 & r> \sigma\,, \end{cases}
\label{eq:pot_hs}
\end{equation}
with $\sigma_s$ playing now the role of the hard-sphere diameter $\sigma$.  
Considering the high-energy deformation penalty $\epsilon$ of micron-sized, non-ionic PNiPAM microgels, the usage of the hard-sphere potential in place of the Hertz potential is well justified (see inset in Fig. \ref{fig:3_g_378}). This is advantageous from a theoretical viewpoint since colloidal hard spheres are among the most thoroughly studied soft matter systems. 

In addition to being the key quantities in static scattering experiments on concentrated suspensions, the static structure factor, $S(q)$, and its associated RDF, $g(r)$, are required as inputs in various methods of calculating colloidal transport properties. For the Hertz potential model, we determine the two pair correlation functions numerically from solving the approximate Percus-Yevick (PY) integral equation \cite{Percus1958,Hansen2006}. The corresponding pair correlation functions for the hard-sphere model are determined from the analytic PY solution combined with the Verlet-Weis (VW) correction  \cite{Verlet1972} incorporating the 
accurate Carnahan-Starling equation of state (see also \cite{Naegele2004}). The VW correction compensates in particular the 
overestimation by the PY solution of the principal peak height, $S(q_m)$, of the hard-sphere structure factor for volume fractions $\phi = (\pi/6) n \sigma^3 \gtrsim 0.4$. Here, $n$ is the number concentration of particles. 
\begin{figure}[hbtp]
\centering
\includegraphics[width=0.5\textwidth]{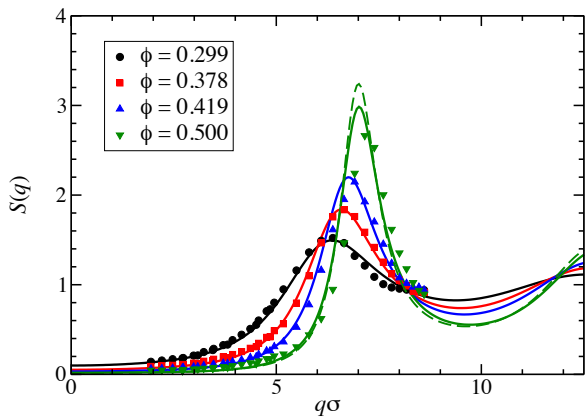}
\caption{Comparison of the experimental static structure factor, $S(q)$, of PNiPAM
 microgels (filled symbols, taken from \cite{Eckert2008}) with the Verlet-Weis corrected Percus Yevick prediction (solid lines), for various particle volume fractions $\phi$ as indicated. The bare PY structure factor for the largest considered volume fraction $\phi=0.5$ is represented by the dashed curve.}
\label{fig:1_struct}
\end{figure}

In Fig. \ref{fig:1_struct}, we compare the PY-VW structre factors for the hard-sphere model    
with static light scattering results by Eckert and Richtering
\cite{Eckert2008} for a concentration series of non-ionic poly(N-isopropylacrylamide) (PNiPAM) microgels in DMF. The details of the particle synthesis are given in \cite{Senff1999a,Stieger2003,Eckert2008}. 
In our PY-VW calculations, we have used the experimentally obtained (mean) 
particle diameter $\sigma = 240\;\text{nm}$, and the (volume-swelling corrected) volume fractions $\phi$ as given in \cite{Eckert2008}. We refer to this reference for the details on how the volume fractions have been determined. The agreement between the experimental and PY-VW structure factors is good even for small wavenumbers $q$. There is in particular a significantly improved agreement for the largest considered volume fraction (dashed curve in Fig. \ref{fig:1_struct}), as compared to the bare PY $S(q)$ used in the earlier work \cite{Eckert2008}. The remaining small deviations for the most concentrated system at $\phi=0.5$  can be at least partially attributed to experimental uncertainties, and  possibly also to the breakdown of the exact factorization of 
the mean scattered intensity into static structure and form factors as discussed by Likos 
\textit{et al.}\cite{Likos2001a,Likos2002}.
\begin{figure}[hbtp]
\centering
\includegraphics[width=0.5\textwidth]{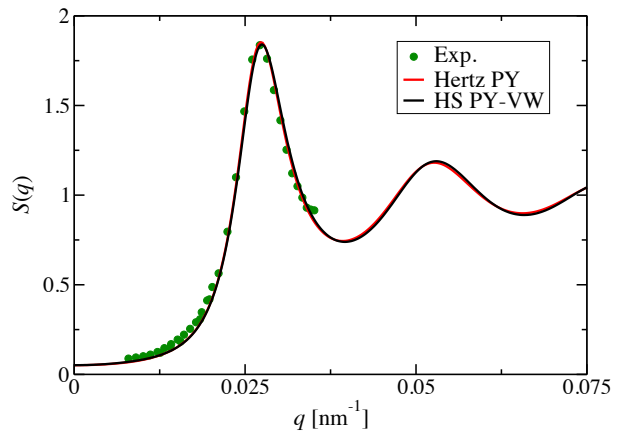}
\caption{Experimental structure factor at $\phi=0.378$  (filled circles) in comparison with the best-fit hard-sphere model PY-VW $S(q)$ (black solid line), and the best-fit Hertz model  $S(q)$ calculated in PY approximation (red solid line). For the Hertz model, the parameters $\epsilon=10^4$ , $\sigma_\text{s}=245\;\text{nm}$, 
and $\phi_\text{s}=0.398$ have been used.}
\label{fig:2_S_378}
\end{figure}
In Fig. \ref{fig:2_S_378}, the experimental structure factor of PNiPAM 
microgels is compared, for $\phi=0.378$, with the best-fit hard-sphere and the best-fit   
Hertz potential structure factors. For the Hertz model system, the parameter values $\sigma_\text{s}=245\;\text{nm}$, $\epsilon=10^4$, and $\phi_\text{s}=(\pi/6)n\sigma_\text{s}^3=0.398$ have been used. The theoretical curves 
coincide practically, as expected for the invoked large value of $\epsilon$ which restricts the softness range of the Hertz potential to a  
narrow interval around the effective diameter $\sigma_\text{s}$. This can be noticed from the inset of Fig. 
\ref{fig:3_g_378}. The remnant potential softness still necessitates a slightly larger effective diameter 
$\sigma_\text{s}$ than the hard-sphere model value $\sigma=240\;\text{nm}$,  
for the Hertz potential incorporates the interactions by dangling polymer chains at the periphery of microgel particles. The larger particle size in the Hertz potential model goes along with a larger volume
fraction $\phi_\text{s}$ than that used in the best-fit hard-sphere model result. With increasing 
volume fraction in the range $\phi_\text{s}\in\left[0.299 -0.5\right]$, a slight decrease of the best-fit effective diameter $\sigma_\text{s}$ is observed. However, owing to the smallness 
of the particle size shrinkage, for the considered PNiPAM in DMF system this effect can be disregarded in its influence on static and dynamic properties. \newline
\begin{figure}[hbtp]
\centering
\includegraphics[width=0.5\textwidth]{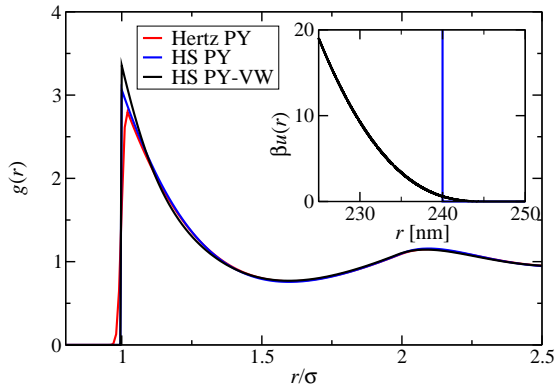}
\caption{Radial distribution functions, $g(r)$, corresponding to the hard-sphere and Hertz potential structure factors in Fig. \ref{fig:2_S_378}. The inset compares the hard-sphere and Hertz pair potential, 
for the parameters in Fig. \ref{fig:2_S_378}.}
\label{fig:3_g_378}
\end{figure}
The $g(r)$'s corresponding to the best-fit  
Hertz and hard-sphere model structure factors in Fig. \ref{fig:2_S_378} are depicted in Fig.  \ref{fig:3_g_378}. The slight softness of the Hertz potential is reflected in the somewhat reduced peak height of $g(r)$ (red solid curve), and in the slight extension of the RDF into the overlap region $r<\sigma_\text{s}$. From Figs. \ref{fig:2_S_378} and \ref{fig:3_g_378}, we conclude that the pair correlations of the considered PNiPAM suspensions in the experimental concentration range are fully compatible with the simple hard-sphere potential. For microgels of lower crosslinker density and smaller size where the softness of the particles is important, the Hertz potential model should be used.

\subsection{Hydrodynamic Particle Modeling}
\label{subsec:hyd_structures}

As an illustration of the concept of a hydrodynamic radius $a_h$ and its associated slip length $L_h$, consider two simple hydrodynamic particle models, namely uniformly fluid-permeable rigid spheres \cite{Felderhof1975b,Felderhof1975}, and non-permeable rigid spheres with Navier partial slip hydrodynamic surface boundary condition \cite{Navier1827}. The physical (material) sphere radius in both models is denoted here by $a$.

In the uniformly-permeable sphere model, the pore-size averaged fluid flow inside a particle is described by the Brinkman-Debye-Bueche (BDB) equation, and the outside flow by the low-Reynolds-number Stokes equation \cite{Kim1991}. The imposed boundary condition is here that the fluid velocity and tangential stress are continuous across the particle surface. The suspension transport properties in this model depend on the material-specific parameter
\begin{equation} \label{eq:permeability}
 \lambda_x = \frac{1}{\kappa\:\!a} \,,
\end{equation}
equal to the ratio of the hydrodynamic penetration length, $1/\kappa$, and particle radius $a$. The penetration length is roughly equal to the mean pore size of the rigid skeleton of a permeable sphere. In the limit $\lambda_x \to 0$ of vanishing mean pore size, a non-permeable sphere with no-slip hydrodynamic boundary condition (BC) on its surface is obtained. For the BDB equation to apply, the mean pore size should be no larger than one tenth of the particle radius so that $\lambda_x \leq 0.1$. The present model has been generalized to spherical particles  with a permeability profile $\kappa(r)$ varying with the radial distance (see, e.g. \cite{Felderhof1975b}) such as core-shell particles
\cite{Masliyah1987a,Zackrisson2006,Cichocki2009,Abade2012} consisting of a dry core and a permeable outer shell. The core-shell model describes in a coarse-grained manner the hydrodynamic effect, e.g., of a polymer brush surrounding the core.

The Navier partial-slip model on the other hand describes fluid-impermeable colloidal spheres where the fluid is allowed to partially slip along their surfaces. The associated Navier partial-slip BC demands for a stationary sphere at each surface point the proportionality of surface-tangential fluid velocity, ${\bf u}_{\|}$, and shear-stress, ${\bf t}_{\|}$, according to
\begin{equation} \label{eq:Navier BC}
{\bf u}_{\|} = \frac{l_s}{\eta_0}\;\!{\bf t}_{\|}\,.
\end{equation}
The proportionality constant is given by the ratio of the
so-called Navier length, $l_s$, and the fluid shear viscosity
$\eta_0$. In the limit $l_s^\ast \to 0$, the no-slip BC describing zero surface slip is recovered. Here, $l_s^\ast=l_s/a$ is the reduced Navier length. In the opposite limit $l_s^\ast \to \infty$, the free-surface BC of zero tangential stress is obtained, corresponding to fluid perfectly slipping along the sphere surface in form of local plug flow. The Navier partial-slip BC can serve as an effective description of a hydrophobic particle surface, and of a rigid particle with surface roughness and corrugations \cite{LECOQ2004}. It is also applicable when non-adsorbing (short) polymers are dispersed in the fluid, owing to the formation of a thin clear-fluid depletion layer at the particle surfaces \cite{Lopez2014}.

The hydrodynamic radius is a single-particle property which depends on the intra-particle hydrodynamic structure of the spherical particle, and in principle also on the considered single-particle transport property. It can be defined operationally through the Stokes-Einstein-Debye expressions
\begin{eqnarray}
D_0^t(\alpha) &=& \frac{k_B T}{6\pi\eta_0\;\!a_h^t(\alpha)} \label{eq:SE-translational}\\
D_0^r(\alpha) &=& \frac{k_B T}{8\pi\eta_0\;\!a_h^r(\alpha)^3} \label{eq:SE-rotational}\,,
\end{eqnarray}
for the translational and rotational diffusion coefficients $D_0^t$ and $D_0^r$, respectively, of an isolated spherical particle in an infinite fluid. An additional definition not considered here is based on the intrinsic viscosity, $[\eta]$, of dispersed spheres \cite{Abade2010}. Here, $\alpha$ denotes a set of parameters characterizing the hydrodynamic particle red structure. For the two considered models is $\alpha = \{\lambda_x,l_s^\ast\}$. The respective translational and rotational hydrodynamic radii are \cite{Felderhof1975b,Felderhof1975}
\begin{eqnarray}
\frac{a_h^t(\lambda_x)}{a} &=& \frac{2x^2\left(x-\tanh\left(x\right)\right)}{2x^3+3(x-\tanh\left(x\right))} \label{eq:hydradius-perm-trans}\\
\frac{a_h^r(\lambda_x)}{a} &=& \left[1+\frac{3}{x^2}-\frac{3\coth\left(x\right)}{x}\right]^\frac{1}{3} \label{eq:hydradius-perm-rot}\,,
\end{eqnarray}
for a homogeneously permeable sphere with $x=1/\lambda_x$, and \cite{Navier1827}
\begin{eqnarray}
\frac{a_h^t(l_s^\ast)}{a} &=& \frac{1 + 2\;\!l_s^\ast}{1 + 3\;\!l_s^\ast} \label{eq:hydradius-Navier-trans}\\
\frac{a_h^r(l_s^\ast)}{a} &=& \left(\frac{1}{1 + 3\;\!l_s^\ast}\right)^{1/3} \label{eq:hydradius-Navier-rot}\,,
\end{eqnarray}
for a Navier partial-slip sphere.

One can associate each hydrodynamic radius $a_h$ with a reduced hydrodynamic slip length, $L_h^\ast$, through 
\begin{equation}
L_h^\ast(\alpha) = \frac{L_h(\alpha)}{a}= 1 -\frac{a_h(\alpha)}{a} \,.
\end{equation}
The quantity $L_h^\ast$ is the relative radial distance, $L_h = a-a_h$, of the apparent no-slip spherical surface inside the particle to its outside surface, in units of the outside surface radius $a$. Except for $\lambda_x$, the asterisk is used throughout to label lengths given in units of $a$, .

In the experimentally common situation where $L_h^\ast$ is small, curvature effects are negligible and the particle-fluid interface can be described as flat. As it is explained in \cite{Cichocki2014}, the key point to notice is that the reduced slip length, $L_{h,f}^\ast=1 -a_{h,f}/a$, in the flat-interface approximation and its associated flat-interface hydrodynamic radius, $a_{h,f}$, are independent of the single-particle transport properties $D_0^t$, $D_0^r$ and $[\eta]$ used in their definition. Since each of these transport properties is associated with a particular ambient velocity field, e.g. with a constant flow field in case of $D_0^t$, an equivalent statement is that $L_{h,f}$ is independent of the ambient flow.

It was shown in \cite{Cichocki2014} that the relations
\begin{eqnarray} 
L_h^\ast(\alpha) &=& L_{h,f}^\ast(\alpha) \left[1 + {\cal O}\left(L_{h,f}^\ast(\alpha)\right) \right] \label{eq:lh-expand}\\
a_h^\ast(\alpha) &=& a_{h,f}^\ast(\alpha) \left[1 + {\cal O}\left(L_{h,f}^\ast(\alpha)^2\right) \right]\label{eq:ah-expand}  \,,
\end{eqnarray}
apply universally to rigid spheres with arbitrary, non-singular spherically symmetric hydrodynamic structures and boundary conditions, independent of the invoked single-particle transport property.

These relations are readily verified, using Eqs. (\ref{eq:hydradius-perm-trans}) - (\ref{eq:hydradius-Navier-rot}), for the special case of a Navier partial-slip sphere where the flat-interface slip length is equal to
\begin{eqnarray}
L_{h,f}^\ast(l_s^\ast) = l_s^\ast  \,,
\end{eqnarray}
and for a uniformly permeable sphere where
\begin{eqnarray}
L_{h,f}^\ast(\lambda_x) = \lambda_x  \,.
\end{eqnarray}
Note that the slip length $L_h$ is equal to the Navier length $l_s$ in the flat-interface limit only.

Eqs. (\ref{eq:lh-expand}-\ref{eq:ah-expand}) do not apply to non-rigid particles such as spherical liquid droplets, and  rigid particles of singular hydrodynamic structure. The latter case is exemplified in \cite{Felderhof1975} by a hollow sphere with a uniformly permeable rigid shell of infinitesimal thickness. The hydrodynamic radii $a_h^t$ and $a_h^r$ of the ultra-thin hollow sphere differ already to linear order in the smallness parameter $\lambda_x$. As for a droplet, an inscribed apparent no-slip spherical surface which is necessarily rigid can not be introduced for the singular hollow-sphere example.

Eqs. (\ref{eq:hydradius-perm-trans}-\ref{eq:hydradius-Navier-rot}) can be inverted 
to obtain the material-specific parameters $\lambda_x$ and $l_s^\ast$ in terms of the ratio
\begin{eqnarray}
\gamma = \frac{a_h}{a} \,,
\end{eqnarray}
or likewise in terms of the reduced width (slip length), $\overline{\gamma}=1-\gamma$, of the apparent fluid annulus region surrounding the apparent no-slip sphere. The parameter $\gamma$ completely characterizes the intra-particle hydrodynamic structure in the hydrodynamic radius model (HRM) where spherical particles of arbitrary hydrodynamic structure are described hydrodynamically as no-slip spheres of reduced radius $a_h$, for unchanged pair potential. This implies in particular an unchanged excluded volume radius $a > a_h$.

\begin{figure}[hbtp]
\centering
\includegraphics[width=0.5\textwidth]{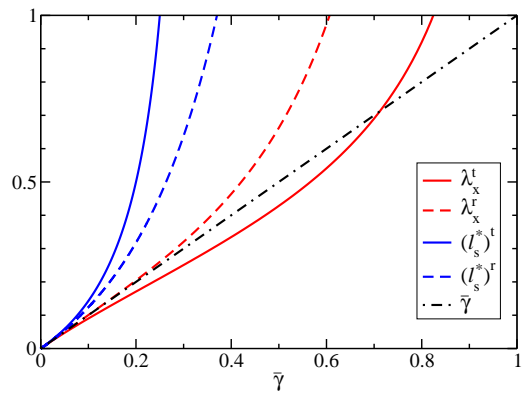}
\caption{Reduced fluid penetration length $\lambda_x^{t,r}$ (red lines) and reduced Navier length $\left(l_s^\ast\right)^{t,r}$ (blue lines), as functions of the reduced fluid annulus width $\overline{\gamma}=1-\gamma$. Solid lines (dashed lines) describe quantities derived from the translational (rotational) single-sphere Stokes-Einstein-(Debye) relation.}
\label{fig:4_l_slip}
\end{figure}

Numerical results illustrating this inversion are depicted in Fig. \ref{fig:4_l_slip} where the reduced penetration length of a permeable sphere, $\lambda_x^{t,r}$, and the reduced slip length of a Navier partial-slip sphere, $\left(l_s^\ast\right)^{t,r}$, are plotted as functions of $\overline{\gamma}$. The reduced lengths derived from the translational (rotational) Stokes-Einstein-(Debye) relation are labeled by the superscript t (superscript r).
For a thin annulus shell with $\gamma > 0.9$, the four reduced lengths are small and commonly  described by the dashed-dotted line $\overline{\gamma}$, corresponding to $\lambda_x\approx L_h$ and $l_s^\ast\approx L_h$. This situation is common to  colloidal systems for which values $\lambda_x < 0.05$ are typically found, and for which the Navier length is commonly equal to a few nanometers.

Differences between the considered lengths are significant for larger values of $\overline{\gamma}$. For a given reduced hydrodynamic radius $\gamma=a_h/a$, two different values $\lambda_x^t \neq \lambda_x^r$ are then obtained. The perfect-slip limit $l_s^\ast \to \infty$ corresponds to $\overline{\gamma}=1/3$ for translational, and to $\overline{\gamma}=1$ for rotational diffusion. While a rotating perfect-slip sphere experiences no friction, akin to a point particle for which $\gamma=0$, the values of $\overline{\gamma}$ for a translating Navier sphere do not exceed $1/3$. Regarding the translational diffusion coefficient, a perfect-slip sphere with $\gamma=2/3$ has the same value, $D_0^t = k_B T/(4\pi\eta_0 a)$, as a uniformly permeable sphere with $\lambda_x \approx 0.278$. However, such a large value of $\lambda_x$ is unrealistic. In order for the BDB equation to describe flow inside a permeable particle, $\lambda_x$ should be no larger than $0.1$, corresponding to a maximally allowed penetration length equal to one-tenth of the particle radius.

\section{Short-time Transport Properties}
\label{sec:short_time_dynamics}

In studying the dynamics in dispersions of colloidal particles, one distinguishes the short-time regime where $\tau_B \ll t \ll \tau_D$ from the long-time regime where $t \gg \tau_D$. Here, $t$ denotes the correlation time of particle concentration fluctuations, and $\tau_B$ and $\tau_D$ are the characteristic particle momentum relaxation and diffusion times, respectively. The time $\tau_B$ is the decay time of velocity correlations of a particle. For submicron sized particles in a low-viscosity fluid such as the here considered PNiPAM in DMF microgels, $\tau_B$ is of the order of microseconds. The particle diffusion time can be estimated by $\tau_D \sim a^2/D_0^t$ which for the considered microgels  is in the millisecond range. In the short-time regime, the particle configuration has changed so little that the slowing influence of non-hydrodynamic direct interactions is not yet operative, different from the solvent-mediated HIs which act quasi-instantaneously on the colloidal scale. Short-time transport properties are thus expressible as simple equilibrium averages, with the direct interactions entering only by their influence on the equilibrium microstructure, e.g on the RDF  $g(r)$. Long-time transport properties are additionally influenced by the Brownian motion of the particles so that they depend in a direct way both on direct and hydrodynamic interactions. 

In this section, we describe our analytic toolbox of calculating short-time diffusion and rheological properties of hydrodynamically structured particles, in the framework of the HRM. In the subsequent sections, the toolbox is used 
for analyzing light scattering data on PNiPAM in DMF suspensions.   

\subsection{Hydrodynamic function} \label{subsec:hydfunction}

The colloidal short-time regime is probed in DLS experiments by analyzing the initially exponential decay,
\begin{equation}
 S(q,t\ll\tau_D) = S(q) \exp\left[-q^2 D\left(q\right)t\right] \,,
\end{equation}
of the dynamic structure factor, $S(q,t)$, in its dependence on the modulus $q$ 
of the single-scattering wave vector ${\bf q}$. The dynamic structure factor is the $q$-th Fourier component of spatio-temporal correlations in thermally induced particle concentration fluctuations. On basis of the many-particle generalized Smoluchowski equation (GSmE), the  wavenumber-dependent diffusion function, $D(q)$, quantifying the short-time exponential decay can be expressed by the ratio \cite{Heinen2011} 
\begin{equation}
D(q) = D_0^t\;\!\frac{H(q)}{S(q)} \,,
\label{eq:diff_func}
\end{equation}
of the hydrodynamic function $H(q)$ and the static structure factor $S(q)$. The hydrodynamic function is the key quantity containing information on colloidal short-time diffusion processes. It can be expressed by the equilibrium configurational average \cite{Nagele1996}, 

\begin{equation}
 H\left(q\right) = \left\langle \frac{k_B T}{N\;\!D_0^t} \sum_{i,j=1}^N \mathbf{\hat{q}} \cdot \bm{\mu}_{ij} \left(\mathbf{X}\right) \cdot \mathbf{\hat{q}}\; e^{i\mathbf{q}\cdot\left[\mathbf{R}_i-\mathbf{R}_j\right]}\right\rangle \,,
\label{eq:hyd_func_1}
\end{equation}
over an ensemble of $N$ colloidal spheres, with the thermodynamic limit $N\to\infty$ for a fixed particle concentration $n$ performed subsequently, in order to describe a macroscopically large scattering volume. Here, 
$\mathbf{\hat{q}}$ is the unit vector in the direction of ${\bf q}$, and $k_B T$ is the system temperature $T$ multiplied by the Boltzmann constant. The translational mobility tensor 
$\bm{\mu}_{ij}$ linearly relates the hydrodynamic force acting on   
particle $j$ at the instant position ${\bf R}_j$ to the resulting velocity change of a particle $i$. It depends on the hydrodynamic structure of the globular particles, e.g. on particle sizes, surface BCs and fluid permeability, and on the many-particles HIs for an instant configuration, ${\bf X}=\{{\bf R}_1,\cdots,{\bf R}_N\}$, of particle centers.

Without HIs, $H(q)$ would be identically equal to one independent of $q$. Any wavenumber dependence reflects thus the presence of HIs. The hydrodynamic function consists of a $q$-independent self-part and a distinct part, i.e.  
\begin{equation}
  H\left(q\right) = \frac{D_S}{D_0^t}  + H_d(q) \,, 
\label{eq:hyd_func_2}
\end{equation}
with the distinct part, $H_d(q)$, approaching its asymptotic value zero for $q \to \infty$. Thus, for large $q$  corresponding to small distances resolved, $H(q)$ reduces to the short-time self-diffusion coefficient, $D_S$, expressed in units of the single-sphere translational diffusion coefficient $D_0^t$. The coefficient $D_S$ quantifies the initial (i.e., short-time) slope of the mean-squared displacement of a particle in presence of other ones. Owing to the slowing influence of the HIs at non-zero particle concentrations, $D_S$ is smaller than $D_0^t$ to which it reduces at infinite dilution only.

The hydrodynamic function has the physical interpretation of a short-time generalized sedimentation coefficient for a  homogeneous suspension of monodisperse colloidal spheres subjected to a weak force field colinear with ${\bf q}$ and oscillating spatially as $\cos\left({\bf q}\cdot{\bf r}\right)$ with positions ${\bf r}$  \cite{Nagele2013b}. For $q \to 0$, a homogeneous force field is recovered such as the local gravitational field on earth.  Consequently,
\begin{equation}
 K  = \frac{V_\text{sed}}{V_0} = \lim_{q \to 0} H(q)
\end{equation}
is the short-time average sedimentation velocity, $V_\text{sed}$, of hydrodynamically interacting colloidal particles, normalized by the particle model dependent mean velocity, $V_0$, of an isolated particle sedimenting in the same force field. The sedimentation coefficient $K$ is related to the short-time collective diffusion coefficient, $D_C$, by
\begin{equation}\label{eq:dcoll}
D_\text{C} = D_0^t\;\!\frac{K}{S(0)}\,,
\end{equation}
where for monodisperse particles $S(0) \equiv lim_{q \to 0} S(q)$ is the relative osmotic compressibility of the  suspension. At larger concentrations where $S(0)$ is small, $D_C$ is significantly larger than $D_0^t$. In principle, the short-time coefficient $D_C$ should be distinguished from the associated long-time collective diffusion coefficient appearing in Fick's law of macroscopic gradient diffusion. The latter, however, is only slightly smaller than the short-time coefficient, by less than $7\;\!\%$ even for a concentrated suspension of no-slip colloidal hard spheres at volume fraction $\phi=0.45$ \cite{Wajnryb2004}. The difference between the two  collective diffusion coefficients can be expected to be even smaller for particles with weaker HIs such as permeable and partial-slip spheres, and charge-stabilized particles with long-range electrostatic repulsion. Different from collective diffusion and sedimentation, the long-time translational self-diffusion coefficient $D_L$ can be substantially smaller than its short-time sibling $D_S$, owing to the retarding relaxation of next-neighbor particle cages which are slightly distorted from their equilibrium spherical symmetry at long times. For colloidal hard spheres at the freezing transition concentration, e.g., $D_L/D_S \approx 0.1$ according to the empirical freezing rule by L{\"o}wen {\em et al.}\cite{Lowen1993a,Nagele2002}.  

In this work, we discuss useful analytic scaling relations for the calculation of short-time diffusion and viscosity properties of suspensions of internally structured particles. These relations are complemented by two semi-analytic and easy-to-implement approximation schemes which have been successfully applied in the past to suspensions of no-slip neutral and charge-stabilized spherical particles \cite{Genz1991,Banchio2008,Heinen2012,Heinen2011}. 
The first scheme is the so-called pairwise-additivity (PA) approximation where all two-body contributions to the hydrodynamic mobilities ${\bm \mu}_{ij}$ are accounted for including near-contact lubrication terms, but three-body and higher-order contributions are disregarded. The PA method is thus well suited for low concentrations where it is basically exact. The second method is a self-part corrected improved version \cite{Heinen2011} of the $\delta\gamma$ renormalized concentration fluctuation expansion by Beenakker and Mazur \cite{Beenakker1984a,Beenakker1984b,Beenakker1983}, where many-particle HIs contributions are approximately accounted for in terms of so-called ring diagrams, but with lubrication corrections disregarded (see also \cite{Makuch2012}). The $\delta\gamma$ method is well suited in particular for larger concentrations. The only input required by these methods is the RDF or likewise the static structure factor. The explicit expressions by both methods for $H(q)$ and the high-frequency (short-time) suspension viscosity $\eta_\infty$, generalized to the HRM, are presented in the Appendices \ref{app:pairwise_additive_approx} and \ref{app:self-part-corrected-Beenakker-Mazur-method}, respectively.  

A key observation regarding the translational mobility tensors ${\bm \mu}_{ij}$ of hydrodynamically structured spherical particles is the relation  
\begin{equation} \label{eq:muexpansion}
  \bm{\mu}_{ij}({\bf X}) = \bm{\mu}_{ij;\textrm{HRM}}({\bf X};a_{h,f}) \left[1 + {\cal O}{(L_{h,f}^\ast}^2) \right]\,,
\end{equation}
where $\bm{\mu}_{ij;\text{HRM}}({\bf X};a_{h,f})$ are the mobility tensors of the associated HRM. 
The relation follows from Eqs. (\ref{eq:lh-expand}) and (\ref{eq:ah-expand}) in conjunction with a general scattering series expansion of the exact $N$-sphere translational mobility tensors \cite{Cichocki2014,Cichocki2013}. Since the short-time transport properties are equilibrium averages of specific mobility tensor elements, it follows that  
\begin{equation}\label{eq:HRM_Hofq_estimate}
  H(q) = H_\textrm{HRM}(q) \left[1 + {\cal O}{(L_{h,f}^\ast}^2) \right]
\end{equation}
with an analogous expression valid for $\eta_\infty$. Thus, the error introduced in calculating short-time transport properties of hydrodynamically structured particles using the simplifying HRM is of ${\cal O}{(L_{h,f}^\ast}^2)$ small. 

\subsection{Translational self-diffusion}
\label{subsec:self-diffusion}

In \cite{Abade2011,Heinen2011}, it was shown  by comparison with computer simulation results for uniformly permeable hard spheres with $\lambda_x < 0.2$ that $D_S$ normalized by its infinite dilution $D_0^t$ is well represented by the simple scaling relation,       

\begin{equation}
\frac{D_\text{S}\left(\phi,\lambda_\text{t}\right)}{D_0^t\left(\lambda_\text{t}\right)} = 1+\lambda_\text{t}\;\!\phi\left(1+0.12\phi-0.70\phi^2\right)\,,
\label{eq:self_diff_improved}
\end{equation}
for all $\phi \leq 0.5$ and with an accuracy of $3.5\;\!\%$ or better. 
A similar scaling relation has been obtained for the short-time rotational self-diffusion coefficient of permeable-sphere suspensions with hard-core interactions \cite{Abade2011}. 
The only dependence in Eq. (\ref{eq:self_diff_improved}) on the permeability parameter $\lambda_x$ characterizing the intra-particle hydrodynamic structure is contained in the second virial coefficient, $\lambda_\textrm{t}=\lambda_t(\lambda_x)$,  which is the linear coefficient in the expansion of $D_S$ in powers of $\phi$. Eq. (\ref{eq:self_diff_improved}) states that the $D_S$ for hydrodynamically structured particles can be scaled to the corresponding coefficient of no-slip hard spheres where $\lambda_x=0$ and $\lambda_t(\lambda_x=0)= -1.8315$.

In fact, on the level of the HRM model Eq. (\ref{eq:self_diff_improved}) is applicable to suspensions of hard spheres of arbitrary hydrodynamic structure, with deviations from the $D_S$ of the actual particle system being of quadratic order small in the reduced slip length $L_h^\ast$ according to Eq. (\ref{eq:HRM_Hofq_estimate}). On recalling that for spherical particles with hard-core direct interactions only the HRM reduces to the spherical annulus model, the only input required in Eq. (\ref{eq:self_diff_improved}) is the second virial coefficient, $\lambda_t(\gamma)$, of spherical annulus particles depending on the ratio, $\gamma=a_h/a$, of  hydrodynamic and hard-core radius. Using for 
$\lambda_t(\gamma)$ the Eq. (\ref{eq:sa_virial_exact}) in Appendix \ref{app:pairwise_additive_approx} in conjunction with tabulated numerical values for the longitudinal and transversal mobility coefficients, $x_{11}(r)$ and $y_{11}(r)$, of two no-slip spheres of radius $a_h$ calculated using the method of Jeffrey and 
Onishi \cite{Jeffrey1984}, we have obtained numerically precise values for the second virial coefficient of spherical annulus particles. These values are described to high accuracy by the polynomial fit     
\begin{align} \label{eq:sa_virial_fit}
\lambda_\text{t}\left(\gamma\right) = &-1.8315 + 7.820 \overline{\gamma} - 14.231 \overline{\gamma}^2   \\
 &+14.908 \overline{\gamma}^3 - 9.383 \overline{\gamma}^4 + 2.717 \overline{\gamma}^5  \,, \notag
\end{align}
accounting for the numerically correct limiting value $\lambda_\text{t}\left(\gamma=1\right)=-1.8315$ of no-slip hard spheres.

\begin{figure}[hbtp]
\centering
\includegraphics[width=0.5\textwidth]{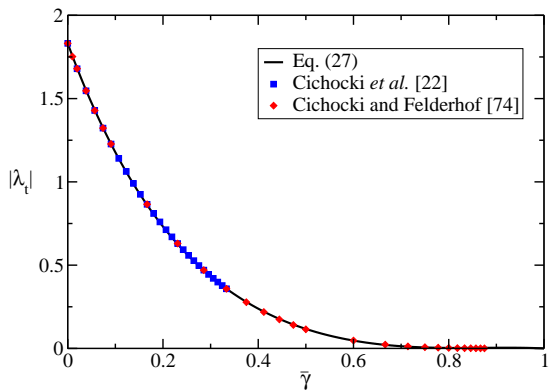}
\caption{Modulus $|\lambda_\text{t}|$ of the first-order virial coefficient of $D_S$ in the spherical annulus model, as a function of the reduced fluid annulus shell width $\overline{\gamma}=1-\gamma$. Solid line: Polynomial fit according to Eq. (\ref{eq:sa_virial_fit}). Filled squares: Tabulated values by Cichocki {\em et al.}  \cite{Cichocki2013} for spherical annulus particles with $\overline{\gamma} \leq 1/3$. Filled diamonds: Tabulated values by Cichocki and Felderhof \cite{Cichocki1991}.}
\label{fig:5_virial_sa}
\end{figure}

In Fig. \ref{fig:5_virial_sa}, the high accuracy of the polynomial fit in Eq. (\ref{eq:sa_virial_fit})  
is shown in comparison with earlier numerical results \cite{Cichocki1991,Cichocki2013} for the second virial coefficient of spherical annulus particles. Note that $\lambda_\text{t}\left(\gamma=0\right)=0$ 
relates to the limiting case of spherical annulus particles interacting hydrodynamically as point particles for which there are no hydrodynamic self-reflections so that $D_S$ reduces to $D_0^t$ independent of $\phi$. The associated long-time coefficient $D_L$ for $\gamma=0$ is still $\phi$-dependent and smaller than $D_0^t$.  

\begin{figure}[hbtp]
\centering
\includegraphics[width=0.5\textwidth]{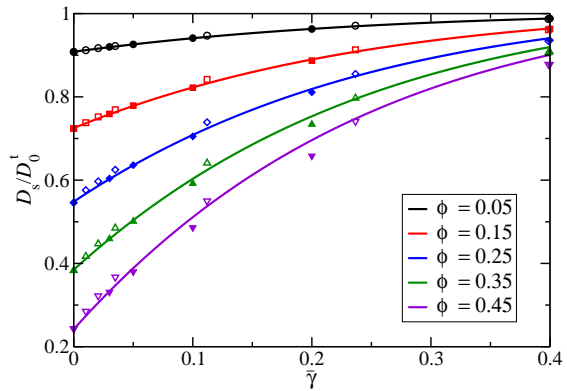}
\caption{Results for the reduced self-diffusion coefficient, $D_S/D_0^t$, of the spherical annulus model as 
function of $\overline{\gamma}=1-\gamma$, for several volume fractions $\phi$ as indicated. Solid lines: Prediction by the scaling formula in Eq. (\ref{fig:5_virial_sa}) in combination with the first order virial coefficient fitting polynomial in Eq. (\ref{eq:sa_virial_fit}). Closed symbols: HYDROMULTIPOLE simulation data for the spherical annulus model tabulated in \cite{Abade2012}. Open symbols: Simulation data for uniformly permeable spheres tabulated in \cite{Abade2010c}, with Eq. (\ref{eq:hydradius-perm-trans}) used in mapping the permeability parameter $\lambda_x$ onto the annulus model parameter $\gamma=a_h/a$ (recall Fig. \ref{fig:4_l_slip}).}
\label{fig:self_diff}
\end{figure}

The accuracy of the analytic scaling formula in  
Eq. (\ref{eq:self_diff_improved}) in combination with Eq. (\ref{eq:sa_virial_fit}) for the short-time self-diffusion coefficient of  spherical annulus particle systems is established in Fig. \ref{fig:self_diff} by the comparison with high-precision 
benchmark simulation data for the spherical annulus \cite{Abade2012} and uniformly 
permeable particle models \cite{Abade2010c}. Regarding the latter model, the conversion of $\lambda_x$ to the related reduced hydrodynamic radius parameter $\gamma$ in the spherical annulus model was done using Eq. (\ref{eq:hydradius-perm-trans}) for $D_0^t(\lambda_x)$. This corresponds to the inversion of the curve for $\lambda_x^t$ in Fig. (\ref{fig:4_l_slip}) in  terms of $\gamma$. For example, the smallest reasonably selected value $\lambda_x=0.1$ corresponds to $\gamma=0.89$.  
Results for $D_S$ are depicted 
in Fig. \ref{fig:self_diff} in dependence on the reduced slip length $\overline{\gamma}$, 
for volume fractions extending over a broad concentration range.   

The excellent agreement between the scaling formula for $D_S$ (solid lines) and the simulation  
data (symbols) does not only validate this formula. For the special case of permeable hard spheres, it additionally   highlights the good performance of the HRM, for a broad concentration range and  
permeability values largely exceeding the ones discussed earlier in the thin shell-limit discussion of the core-shell model in \cite{Cichocki2013}.

In summary, the analytic formula in Eqs. (\ref{eq:self_diff_improved}) and (\ref{eq:sa_virial_fit}) allows for a quick and accurate calculation of the translational short-time self-diffusion coefficient of hydrodynamically structured spherical particles with hard-core interactions. The only input is the single-particle property $a_h/a$ which can be determined experimentally, e.g., by a DLS measurement of $D_0^t$ in conjunction with a static scattering experiment determining the excluded volume radius. The formula for $D_S$ is applicable also to particles with a short-range, weakly soft pair potential such as the Hertz potential for non-small potential strengths $\epsilon$, provided the effective diameter $\sigma_\text{s}$ and the related volume fraction $\phi_\text{s}$ are used instead of $\sigma$ and $\phi$. 

The formula for $D_S$ in Eqs. (\ref{eq:self_diff_improved}) and (\ref{eq:sa_virial_fit}) based on its virial expansion 
is not valid for particles with a long-range, soft pair potential where an (effective) excluded volume diameter is not the characteristic parameter. An example in case are low-salinity suspensions of charge-stabilized colloidal spheres, where according to theory, simulation and experiment $D_S$ has to good accuracy the fractional concentration dependence $D_S/D_S^0 -1 \approx A_S\;\!\phi^{4/3}$, with a coefficient $A_S \eqsim 2.5 - 2.9$ 
which varies to a small extent with the particle size and charge \cite{Banchio2008,Heinen2010}. The initial slope of $D_S$ at $\phi=0$ is thus zero in these charge-stabilized systems. Different from the formula in Eqs. (\ref{eq:self_diff_improved}) and (\ref{eq:sa_virial_fit}), the HRM is applicable also to spherical particles with arbitrary soft direct interactions, and to particles of fuzzy hydrodynamic structure without a sharp outer boundary. In the framework of the HRM, the coefficient $D_S$ and other short-time transport properties of particles with soft interactions can be approximately and semi-analytically calculated using the PA method at smaller and the self-part corrected $\delta \gamma$ method at larger concentrations. The inputs $g(r)$ and $S(q)$ to these methods can be obtained from Ornstein-Zernike integral equation schemes such as the analytical scheme for charge-stabilized 
particles introduced in \cite{Heinen2011c,Heinen2011d}.

\subsection{Sedimentation coefficient}
\label{subsec:sedimenation}

Different from self-diffusion, the neglect of hydrodynamic near-distance and particularly lubrication effects 
in approximate analytic calculations is less consequential regarding sedimentation. Therefore, $K$ can be described semi-qualitatively on the Rotne-Prager (RP) type level where only the long-distance dipolar contribution to the mobility tensors is accounted for. This amounts hydrodynamically to the neglect of all hydrodynamic flow reflection by the particles.

\begin{figure}[hbtp]
\centering
\includegraphics[width=0.5\textwidth]{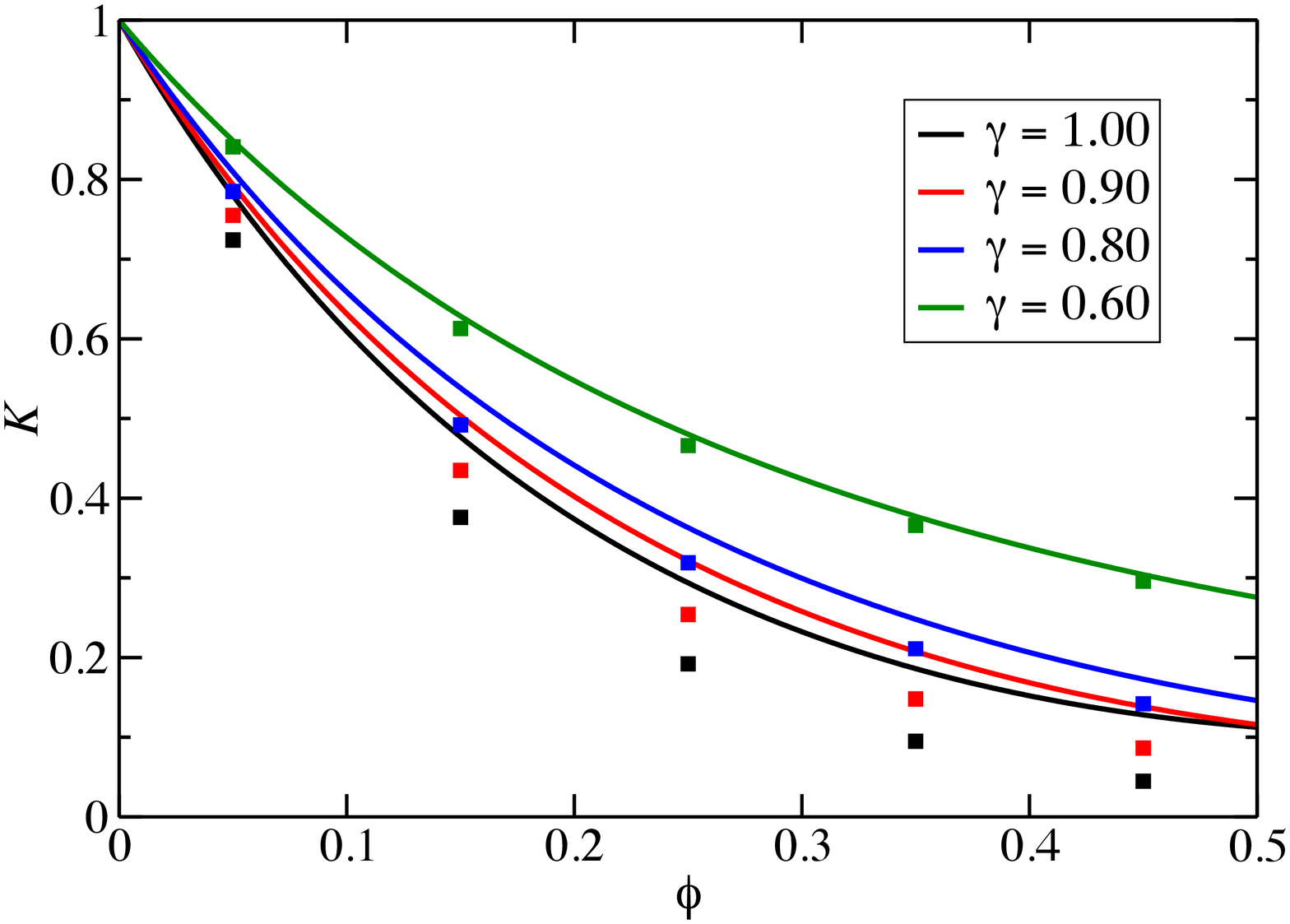}
\caption{Concentration dependence of the sedimentation coefficient, $K$, of the spherical annulus model for reduced hydrodynamic radius values $\gamma$ as indicated. 
Solid lines: RP approximation $K_\textrm{RP}$ in Eq. (\ref{eq:K_annulus}). Filled symbols: Simulation data taken from \cite{Abade2012}.}
\label{fig:8_K_py_rp_vs_sim}
\end{figure}

Using the RP approximation in conjunction with the analytic PY solution for the Laplace transform of $r\;\!g(r)$, Contreras-Aburto \textit{et al.} \cite{Contreras-Aburto2013} have derived analytic expressions for the short-time sedimentation  
coefficient of Navier partial-slip and uniformly permeable spheres with hard-core interaction.   
Here, we present the according expression for spherical annulus particles,
\begin{equation}
 K_\textrm{RP}\left(\phi,\gamma\right) = 1+\gamma\;\!\phi\left(\gamma^2 +12\left[\frac{\phi\left(2-\phi\right)-10} {20\left(1+2\phi\right)}\right]\right)\,,
\label{eq:K_annulus}
\end{equation}
which includes hydrodynamic point particles as a limiting case for which $K(\phi,\gamma=0)=1$. In the opposite limit,
$\gamma=1$, of no-slip hard spheres, Eq. (\ref{eq:K_annulus}) reproduces an expression by Banchio and N\"agele \cite{Banchio2008} which was rederived subsequently by Gilleland {\em et al.} \cite{Gilleland2011} using a variational method. 

The comparison in Fig. \ref{fig:8_K_py_rp_vs_sim} of the RP based analytic formula in Eq. (\ref{eq:K_annulus}) with benchmark simulation data for spherical annulus hard spheres \cite{Abade2012} shows that the sedimentation velocity is overestimated by the formula at larger concentrations. This is a consequence of the neglect of flow reflections 
in the RP approximation which becomes less severe with decreasing annulus parameter $\gamma$, owing to the for 
a fixed $\phi$ increasing distances between the hydrodynamic particle surfaces. For the lowest considered value $\gamma=0.6$, excellent agreement between the simulation data and $K_\textrm{RP}$ is observed. The largest deviations occur for no-slip hard spheres where $K_\textrm{RP}$ provides an upper bound to the exact sedimentation coefficient \cite{Gilleland2011,Nagele2013b}. As an aside, we note that even at $\phi=0.5$, 
$K_\textrm{RP}$ changes only slightly if the VW-corrected $g(r)$ 
is used instead of the bare PY $g(r)$.  
\begin{figure}[hbtp]
\centering
\includegraphics[width=0.5\textwidth]{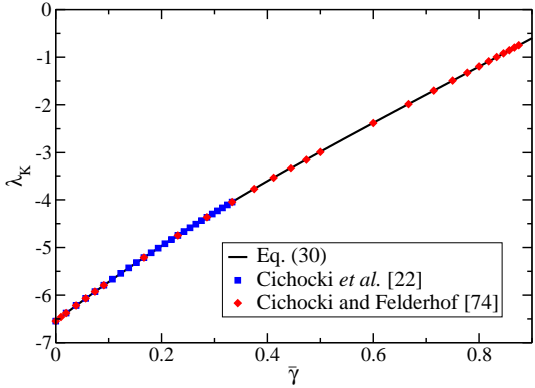}
\caption{First-order virial coefficient, $\lambda_K$, of the sedimentation coefficient of spherical annulus particles.  Filled symbols: Tabulated values by Cichocki {\em et al.} for thick (diamonds) \cite{Cichocki1991} and thin annulus shell (squares) systems \cite{Cichocki2013}, in comparison with the polynomial in Eq. (\ref{eq:K-virialcoeff-fit}) (black solid line).}
\label{fig:7_virial_k}
\end{figure}

While $K_\textrm{RP}$ nicely describes the trends of the exact spherical annulus sedimentation coefficient $K$ 
in its $\phi$ and $\gamma$ dependence, in search of an improved analytic expression we make the ansatz 
\begin{align} \label{eq:K-scaling}
K\left(\phi,\gamma\right) &= 1 + \lambda_\text{k}\left(\gamma\right)\,u_\text{k}\left(\phi,\gamma\right) \\
&=1 + \lambda_\text{k}\left(\gamma\right)\phi\left[1+\mathcal{O}\left(\phi\right)\right]\,, \notag
\end{align}
where $\lambda_\text{K}\left(\gamma\right)$ is the first-order virial 
coefficient of the sedimentation coefficient for which high-precision values have been provided by Cichocki \textit{et al.}  \cite{Cichocki1991,Cichocki2013}. According to Fig. \ref{fig:7_virial_k}, these tabulated values are well represented by the polynomial, 
\begin{align} \label{eq:K-virialcoeff-fit}
  \lambda_\text{K}\left(\gamma\right) = -6.5464 &+8.592  \overline{\gamma}
  -3.901 \overline{\gamma}^2 \\ 
  &+ 2.011 \overline{\gamma}^3 - 0.142 \overline{\gamma}^4  \notag
\end{align}
in the full parameter range $0 <\gamma \leq 1$.  
At $\gamma=1$ in particular, the known value $\lambda_\textrm{K}=-6.5464$ of no-slip hard spheres is recovered from the polynomial. 

In the first equality in Eq. (\ref{eq:K-scaling}), we have introduced the sedimentation scaling function 
$u_\text{K} = \left(K-1\right)/\lambda_\text{K}$. 
In \cite{Abade2011}, a scaling ansatz analogous to Eq. (\ref{eq:K-scaling}) was used for the short-time translational and  rotational self-diffusion coefficients of uniformly permeable hard spheres. By the comparison with simulation data, 
it was shown therein that the scaling functions $u_\textrm{S}(\phi,\lambda_x)$ and $u_\textrm{R}(\phi,\lambda_x)$ associated with translational and rotational self-diffusion, respectively, are practically independent of the permeability coefficient for all values $\lambda_x \leq 0.1$. They are therefore well approximated by (i.e. scaled to) the respective functions $u_\textrm{S,R}(\phi,\lambda_x=0)$ of non-permeable solid spheres. 
On using a third-order polynomial fit of $u_\textrm{S}(\phi,\lambda_x=0)$ obtained from simulation data of no-slip hard spheres, and the first two known virial coefficients of $D_S(\phi,\lambda_x=0)$,  
Eq. (\ref{eq:self_diff_improved}) for $D_S(\phi,\lambda_x)$ has been obtained \cite{Abade2011,Heinen2011}. 
\begin{figure}[hbtp]
\centering
\includegraphics[width=0.5\textwidth]{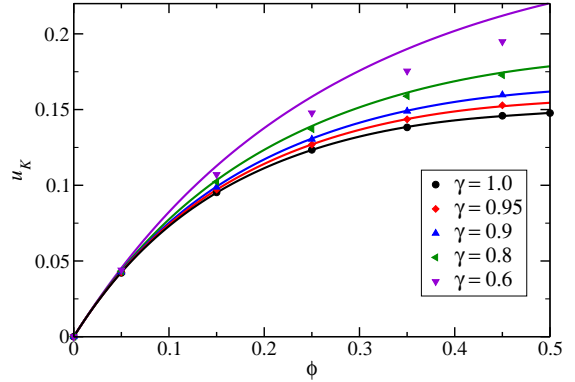}
\caption{Concentration dependence of the scaling function $u_\text{K}\left(\phi,\gamma\right)$ associated with the sedimentation coefficient of spherical annulus particles, for values of $\gamma$ as indicated. Colored closed symbols: Values obtained from simulation data of $K(\phi,\gamma)$ \cite{Abade2012}. 
Colored solid lines: Semi-empirical formula in Eq. (\ref{eq:sed_scal_func_fit}).}
\label{fig:6_K_red_test_scale}
\end{figure}

As noticed already in the context of permeable spheres \cite{Abade2011}, $u_\textrm{K}$   
depends significantly on the intra-particle hydrodynamic structure, different from its self-diffusion siblings. In Fig. \ref{fig:6_K_red_test_scale}, 
this is demonstrated for the spherical annulus system using tabulated simulation data for $K(\phi,\gamma)$ 
in \cite{Abade2012}. 
The simulation data of $u_\textrm{K}(\phi,\gamma)$ for a fixed $\phi >0$ clearly differ from each other 
for different $\gamma$ values. Thus, the $\gamma$-dependence of $K(\phi,\gamma)$ cannot be embodied solely in terms of the first-order virial coefficient, in contrast to Eq. (\ref{eq:self_diff_improved}) describing $D_S$.  
However, as it is shown in Fig. \ref{fig:6_K_red_test_scale}, the simulation data for $u_\textrm{K}$ are well 
described for $\gamma \leq 0.8$ by a forth-order polynomial in $\gamma\phi$, namely  
\begin{align}\label{eq:sed_scal_func_fit}
u_\text{K}\left(\phi,\gamma\right) = \phi &\left[1 -3.348\gamma\phi+7.426\left(\gamma\phi\right)^2 \right. \\
&\left.\quad-10.034\left(\gamma\phi\right)^3+5.882\left(\gamma\phi\right)^4\right] \,. \notag
\end{align}
The resulting analytic expression,
\begin{align}\label{eq:K_fit}
K\left(\phi,\gamma\right) &= 1 -  \, \lambda_\text{K}\left(\gamma\right) \phi \left[ 1 -3.348\gamma\phi\right. \\ 
&\left. +7.426\left(\gamma\phi\right)^2 -10.034\left(\gamma\phi\right)^3+5.882\left(\gamma\phi\right)^4\right] \,, \notag
\end{align}
for the sedimentation coefficient provides in conjunction with Eq. (\ref{eq:K-virialcoeff-fit}) for $\lambda_K(\gamma)$ an accurate description in the from the experimental 
viewpoint sufficiently broad parameter range $\gamma \in \{0.8-1\}$. 
The numerical coefficient $3.348$ in the bracket of Eq. (\ref{eq:K_fit}) is selected 
such that at $\gamma=1$ the correct numerical value $21.918$ of the second virial coefficient 
of no-slip rigid spheres \cite{Cichocki1991} is recovered.\\
\begin{figure}[h]
\centering
\includegraphics[width=0.5\textwidth]{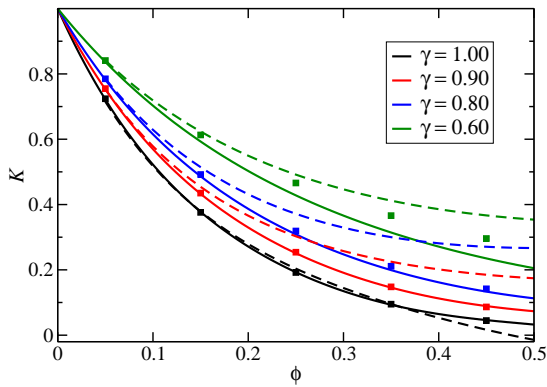}
\caption{Sedimentation coefficient of the spherical annulus model as a function of $\phi$, for values of $\gamma$ as indicated. Filled squares: Simulation data for spherical annulus particles \cite{Abade2012}. Solid lines: Analytic formula in Eq. (\ref{eq:K_fit}) with $\lambda_\textrm{K}$ according to Eq. (\ref{eq:K-virialcoeff-fit}). Dashed lines: Self-part corrected $\delta\gamma$ method prediction, with self-part $D_S$ according to Eq. (\ref{eq:self_diff_improved}) and VW-PY input for $S(q)$.}
\label{fig:9_K_vs_L_h}
\end{figure}

From Fig. \ref{fig:9_K_vs_L_h}, the good agreement of the semi-empirical formula for $K(\phi,\gamma)$ in Eq. (\ref{eq:K_fit}) 
with the spherical annulus simulation data \cite{Abade2012}  
is noticed for $\gamma \geq 0.8$. The figure depicts also simulation data for permeable spheres 
where $\lambda_x$ has been converted to the respective $\gamma$ using Eq. (\ref{eq:hydradius-perm-trans}). 
Moreover, results for $K(\phi,\gamma)$ are displayed as predicted by the  
self-part corrected $\delta\gamma$ scheme with VW-PY structure factor input where the self-part contribution to $K$  was calculated according to Eq. 
(\ref{eq:self_diff_improved}). See Appendix \ref{app:self-part-corrected-Beenakker-Mazur-method} for details on the $\delta\gamma$ method of calculating $H(q)$ and its self-part correction. While in good overall 
agreement with the simulation data, the self-part corrected $\delta\gamma$ method results 
for $K$ deviate significantly at larger $\phi$. Different from Eq. (\ref{eq:K_fit}) which applies to spherical particles with hard-core interactions only, the $\delta\gamma$ method is applicable also to particles with soft interactions.

The analytic expressions in Eqs. (\ref{eq:K_fit}) and (\ref{eq:self_diff_improved}) for $K$ 
and $D_S$ are profitably used in the following discussion of the hydrodynamic 
function of core-shell particle systems.

\subsection{Hydrodynamic function scaling}
\label{subsec:hydrodynamic-function}

Due to the fact that $H(q)$ is given according to Eq. (\ref{eq:hyd_func_1}) 
by an equilibrium average over hydrodynamic mobilities, 
its principal peak location and the wavenumber locations of its secondary maxima are nearly coincident with those of $S(q)$. 
As a static equilibrium property, $S(q)$ is independent of the hydrodynamic particle structure 
and the HIs in general.  
This observation has led Abade {\em et al.} to the following remarkable finding \cite{Abade2010d},  
which they analyzed in the context of uniformly permeable hard spheres: 
While the amplitudes of the oscillations in $H(q)$ are strongly sensitive to the permeability (i.e., the hydrodynamic particle structure), the relative $q$-dependence of $H(q)$ is practically permeability independent 
and can be scaled thus to that of no-slip hard spheres. 
To see this quantitatively, consider the so-called reduced hydrodymanic function \cite{Abade2010d}, 
\begin{equation}
h_d\left(q\right) = \frac{H_d(q)}{\left|H_d(q=0)\right|}\,,
\label{eq:red_func}
\end{equation}
where $H_d(q)$ is the wavenumber-dependent distinct part of $H(q)$ introduced in Eq. (\ref{eq:hyd_func_2}).
\begin{figure}[hbtp]
\centering
\includegraphics[width=0.5\textwidth]{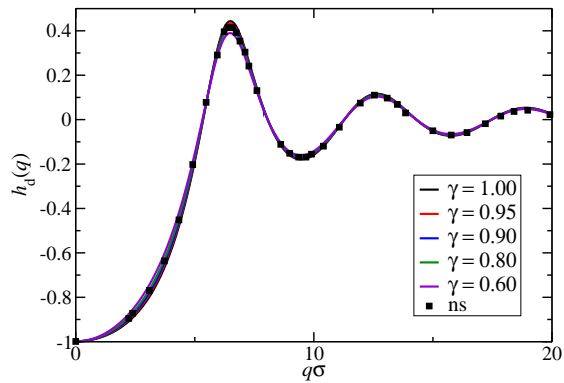}
\caption{Reduced hydrodynamic function, $h_d(q)$, for spherical annulus particles at fixed volume fraction $\phi=0.35$ and varying $\gamma$ as indicated. The wavenumber is scaled in terms of the hard-sphere diameter $\sigma=2a$. Solid lines: $\delta\gamma$ method results using VW-PY structure factors as input (cf. Appendix \ref{app:self-part-corrected-Beenakker-Mazur-method}). Filled squares: Simulation result for no-slip (ns) hard spheres ($\gamma=1$) taken from \cite{Abade2010d}.}
\label{fig:10_h_red_0_35}
\end{figure}

The reduced hydrodynamic function is defined such that $h_d(q=0)=-1$ and $h_d(q\to\infty)=0$. 
Abade {\em et al.} found that $h_d(q)$ is at all $q$ nearly independent of the permeability parameter $\lambda_x$, 
practically in the complete liquid-phase concentration interval. 
In extrapolating their finding to the spherical annulus model as motivated 
by our general discussion in Subsec. \ref{subsec:hyd_structures}, $H(q)$ 
can be expected to be well represented by 
\begin{equation}
 H\left(q\right) \approx \frac{D_\text{S}\left(\phi,\gamma\right)}{D_0^\text{t}\left(\gamma\right)} + h_d^{ns}\left(q\right)
 \left[K(\phi,\gamma)-\frac{D_S\left(\phi,\gamma\right)}{D_0^\text{t}\left(\gamma\right)}\right] \,,
\label{eq:red_hyd_func}
\end{equation}
where $h_d^{ns}(q)=h_d(q,\gamma=1)$ is the reduced hydrodynamic function of no-slip hard spheres. The hydrodynamic particle structure enters into this expression by the coefficients $K$ and $D_S/D_0^\text{t}$ only for which we have provided  accurate analytic expressions. The relative $q$-dependence of $H(q)$ is described by 
the master function $h_d^{ns}(q)$ which can be conveniently calculated 
using the semi-analytic $\delta\gamma$ method for $H_d(q)$. The $\delta\gamma$ method has been shown, 
in comparison to Stokesian dynamics simulations, to give quite accurate predictions 
not only for the $H_d(q)$ of no-slip hard spheres \cite{Heinen2011}, 
but likewise for charge-stabilized suspensions with long-range electrostatic interactions \cite{Heinen2011,Heinen2012}.  
\begin{figure}[hbtp]
\centering
\includegraphics[width=0.5\textwidth]{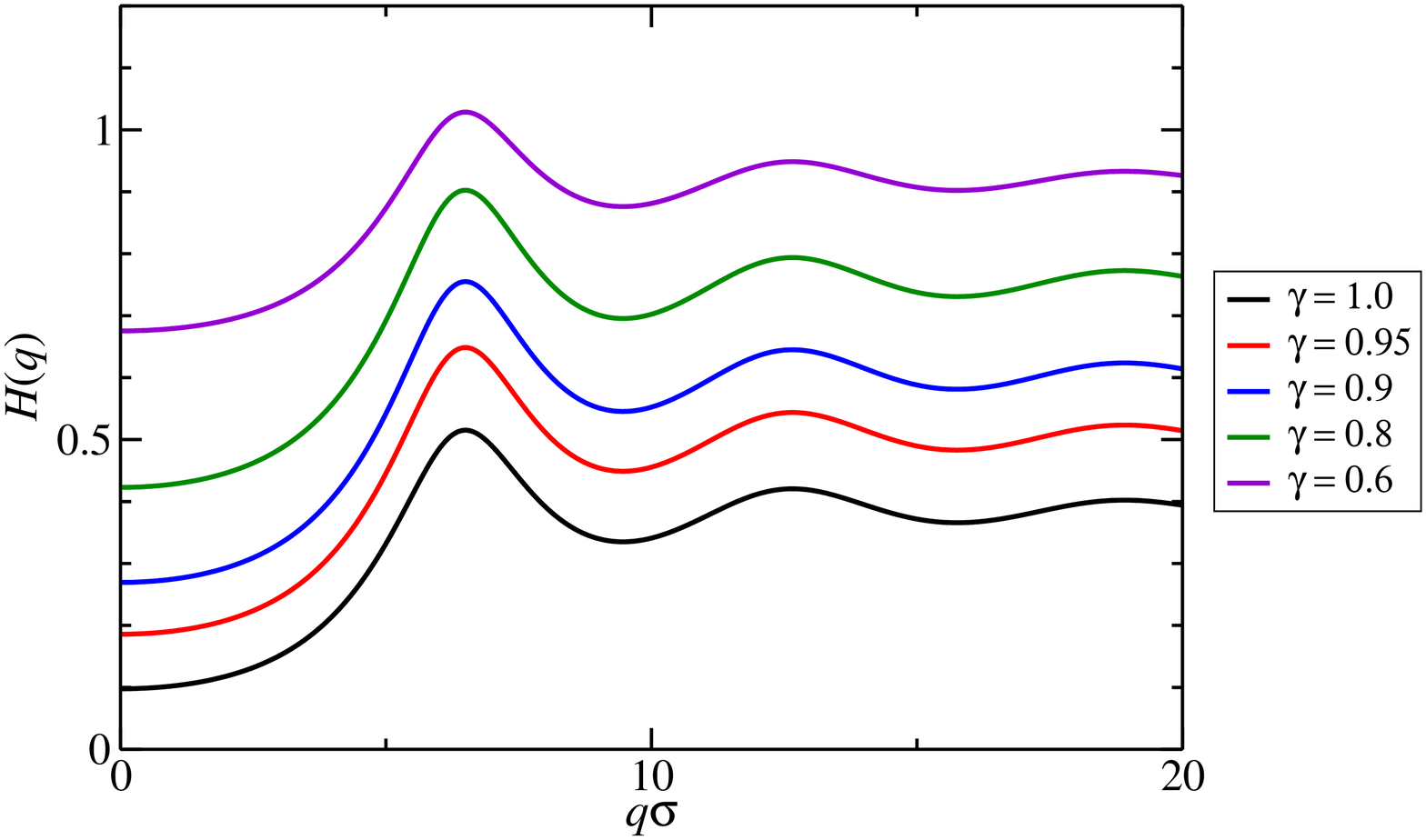}
\caption{Hydrodynamic function, $H(q)$, of spherical annulus spheres for $\phi=0.35$ and values of $\gamma$ as indicated. The solid lines show results by the semi-analytic formula in Eq. (\ref{eq:red_hyd_func}), with $D_S/D_0^\text{t}$ and $K$ according to Eqs. (\ref{eq:self_diff_improved}) and (\ref{eq:K_fit}), respectively, and $h_d^{ns}(q)$ calculated using the $\delta\gamma$ method.}
\label{fig:12_H_by_reduced}
\end{figure}
To validate hydrodynamic function scaling for the spherical annulus model, in Fig. \ref{fig:10_h_red_0_35} 
we present results for $h_d(q,\gamma)$ in a broad $\gamma$ parameter range, obtained 
using the $\delta\gamma$ method in Appendix \ref{app:self-part-corrected-Beenakker-Mazur-method}. 
All curves collapse practically on a single master curve which 
in turn nicely agrees with the simulation data for no-slip (ns) hard spheres taken from \citep{Abade2010d}. 

The sensitivity of $H(q)$ on the reduced hydrodynamic radius $\gamma$ is illustrated in Figure \ref{fig:12_H_by_reduced}, where $H(q)$ has been calculated according to Eq. (\ref{eq:red_hyd_func}). The strength of the HIs ceases with decreasing $\gamma$. Notice that $H(q) \rightarrow 1$ for $\gamma\rightarrow 0$.

\begin{figure}[hbtp]
\centering
\includegraphics[width=0.5\textwidth]{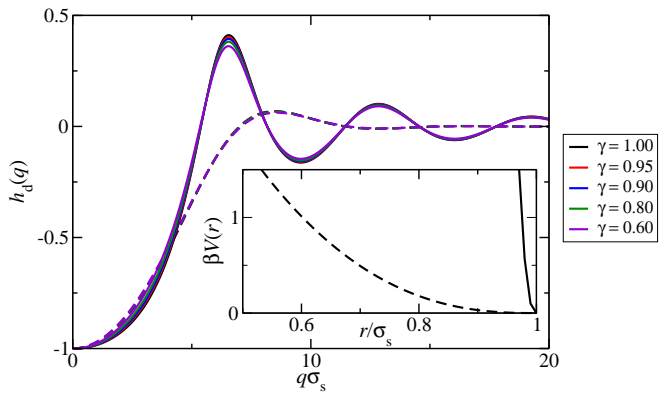}
\caption{Reduced hydrodynamic function, $h_d(q)$, for a Hertzian spheres system of interaction strength $\epsilon=10$ (dashed lines) and $\epsilon=10^4$ (solid lines), respectively, for various reduced hydrodynamic radii, 
$\gamma=a_h/a_s$, as indicated. The effective volume fraction is $\phi_\text{s}=0.35$. The depicted results have been calculated using the $\delta\gamma$ method with PY structure factor input. Inset: Excerpt of the Hertz potential curve for $\epsilon=10$ (dashed) 
and $\epsilon=10^4$ (solid), respectively. Wavenumber $q$ and radial distance $r$ are scaled by the effective soft  
diameter, $\sigma_\text{s}=2 a_s$, of the Hertz potential.}
\label{fig:11_h_red_0_35_hertz_10_10000}
\end{figure}

To date, the validity of the hydrodynamic function scaling was scrutinized 
for particles with hard-sphere interactions only 
\cite{Abade2010d,Abade2010c}. As discussed in 
Subsec. \ref{subec:pair_correlations}, the soft Hertz potential in Eq. (\ref{eq:pot_hertz}) 
is a useful description of the coarse-grained interaction of certain types of mechanically soft 
microgel particles. For this reason, we investigate now the scaling of 
the hydrodynamic function for suspensions of Hertz particles of different interaction 
strengths $\epsilon$. In Fig. 
\ref{fig:11_h_red_0_35_hertz_10_10000}, the functions $h_d(q)$ of Hertz particles are shown for 
various values of the reduced hydrodynamic radius, defined here by $\gamma = a_h/a_s$ with $a_s=\sigma_s/2$ 
denoting the effective soft radius of the Hertz potential. Two  
largely distinct interaction (softness) parameters $\epsilon=10$ and $10^4$ are considered, 
representing highly soft and weakly soft particle systems, respectively. 
The inset depicts the respective shapes of the Hertz potential. For $\epsilon=10$, there is a 
significant likelihood of finding two particles at a distance smaller than $\sigma_\text{s}$, as quantified by 
values of $g(r<\sigma_\text{s})$ significantly larger than zero. Even then the HRM remains applicable, 
provided the hydrodynamic structure of the actual soft particles is not 
significantly distorted away (on average) from spherical symmetry during particle interpenetration.    

The curves for $h_d(q)$ in 
Fig. \ref{fig:11_h_red_0_35_hertz_10_10000} have been obtained using the 
HRM-based $\delta\gamma$ method with the structure factor input for the Hertzian spheres calculated in PY approximation. 
The figure shows that hydrodynamic function scaling applies to Hertz model particles for a broad softness range, with the shape of the $\gamma$-independent master curve for $h_d(q)$ depending 
on the softness parameter. The scaling behavior of $H(q)$ can be expected to 
hold also for other soft pair potentials.  

\subsection{High-frequency viscosity}
\label{subsec:high-frequency-viscosity}

So far diffusion related properties have been addressed only. 
We discuss next a short-time 
rheological property, namely the high-frequency-limiting suspension viscosity, $\eta_\infty$, which 
linearly relates the average suspension shear stress to the applied rate of strain in a low-amplitude, 
high-frequency shear experiment. Like the short-time diffusion properties discussed before, 
$\eta_\infty$ is a quantity of purely hydrodymanic origin, influenced by direct particle interactions through the equilibrium averaging only. For particles acting hydrodynamically like points ($\gamma=0$), 
it reduces thus to the shear viscosity, $\eta_0$, of the suspending Newtonian fluid. 

The high-frequency viscosity should be distinguished from the zero-frequency viscosity \cite{Russel1984},
\begin{equation}
 \eta\left(\phi\right) = \eta_\infty\left(\phi\right)+\Delta\eta\left(\phi\right) \,,
\label{eq:zero_freq}
\end{equation}
determined in a steady low-shear experiment. The zero-frequency viscosity $\eta$ has an additional  
contribution, $\Delta \eta>0$, originating from the relaxation of the shear-distorted dynamic particle cage formed 
around each particle. The shear relaxation part $\Delta\eta$ is influenced both by direct and hydrodynamic interactions, 
with the consequence that for strongly correlated colloidal particles it is substantially larger than $\eta_\infty$. 
An analytic method of calculating the long-time transport property $\eta$ is 
presented in Subsec. \ref{subsec:steady-shear-viscosity}.
\begin{figure}[hbtp]
\centering
\includegraphics[width=0.5\textwidth]{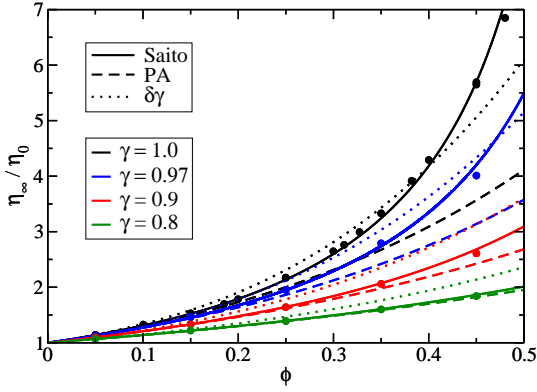}
\caption{High-frequency viscosity, $\eta_\infty$,  of spherical annulus particles in dependence on $\phi$, for values of $\gamma$ as indicated. Filled symbols: Simulation data taken from \cite{Abade2012,Banchio2008}. 
Solid lines: Generalized Sait\^o formula in Eqs. (\ref{eq:Saito_formula}) - (\ref{eq:lambdaVpolynomial}). 
Dashed lines: PA scheme results with VW-PY input for $g(r)$. Dotted lines: $\delta\gamma$ scheme results with VW-PY input for $S(q)$.}
\label{fig:18_vis_high_nu_comp}
\end{figure}

In \cite{Abade2010a,Abade2010d}, a generalized Sait\^o formula for the high-frequency viscosity of permeable spheres with 
hard-core interactions has been introduced which for $\phi \leq 0.5$ gives results in good agreement with simulation data. In the framework of the spherical annulus model, the formula reads   
\begin{equation} 
\frac{\eta_\infty\left(\phi,\gamma\right)}{\eta_0} = 1+\left[\eta\right](\gamma) \phi\frac{1+\hat{S}\left(\phi,\gamma\right)}{1-\frac{2}{5}\left[\eta\right](\gamma)\phi\left(1+\hat{S}\left(\phi,\gamma\right)\right)} \,.
\label{eq:Saito_formula}
\end{equation}
It expresses $\eta_\infty$ in terms of the intrinsic viscosity,  
$\left[\eta\right]\left(\gamma\right) = (5/2)\;\!\gamma^3$, depending on the 
hydrodynamic particle structure only, and the Sait\^o function $\hat{S}\left(\phi,\gamma\right)$. The latter is  approximated linearly in $\phi$ as
\begin{equation}
\hat{S}\left(\phi,\gamma\right) = \left( \frac{\lambda_V(\gamma)}{[\eta](\gamma)}-\frac{2}{5}[\eta](\gamma)\right)
\phi\,,
\end{equation}
where $\lambda_V(\gamma)$ is the first-order virial coefficient in the expansion of $\eta_\infty/\eta_0$ in powers of $\phi$. Numerical values for the first virial coefficient of spherical annulus particles are given in \cite{Cichocki2013}. For $\gamma \geq 2/3$, these values are well represented by the polynomial
\begin{align} \label{eq:lambdaVpolynomial}
 \lambda_\text{V}(\gamma)= 5.0021 &-39.279 \overline{\gamma} +143.179 \overline{\gamma}^2 \\
 &-288.202 \overline{\gamma}^3 + 254.581 \overline{\gamma}^4\,.\notag
\end{align}

In Fig. \ref{fig:18_vis_high_nu_comp}, the predictions for $\eta_\infty$ by the generalized Sait\^o 
formula in Eqs. (\ref{eq:Saito_formula}) - (\ref{eq:lambdaVpolynomial}) are compared 
with existing simulation results \cite{Abade2012} for the spherical annulus model. 
There is good agreement 
with the simulation in the displayed liquid-phase concentration range. For fixed concentration $n$ and fixed hard-core radius $a$, the viscosity increases with increasing $a_h$, i.e. increasing $\gamma$, owing to the enlarged dissipation. 
Additionally shown in the figure   
are results for $\eta_\infty$ by the PA and $\delta\gamma$ methods described in Appendices \ref{app:pairwise_additive_approx} and  
\ref{app:self-part-corrected-Beenakker-Mazur-method}, respectively. 
Like in the PA scheme for 
short-time diffusion properties, two-body HIs contributions to $\eta_\infty$ 
are fully accounted for but three-body and higher order 
contributions have been neglected. The PA scheme is in good agreement with the simulation data 
for $\phi<0.2$ only. For small $\gamma$, the applicability of the PA method extends to somewhat 
larger $\phi$. We attribute this first to the weaker hydrodynamic interactions for $\gamma<1$, and 
second to the fast ${\cal O}(1/r^6)$ asymptotic decay of the 
shear mobility function associated with $\eta_\infty$ (see Appendix \ref{app:pairwise_additive_approx}).
While the $\delta\gamma$ scheme viscosity predictions for no-slip spheres are 
in better agreement with the simulation data than those by the PA scheme, 
for $\gamma<1$ the $\delta\gamma$ scheme  
consistently overestimates the high-
frequency viscosity. Quite interestingly, for all $\phi < 0.5$ the  
$\delta\gamma$ prediction for $\eta_\infty$ scales to good accuracy with the hydrodynamic volume fraction 
$\phi_\text{h}=\gamma^3\phi$. 
\begin{figure}[hbtp]
\centering
\includegraphics[width=0.5\textwidth]{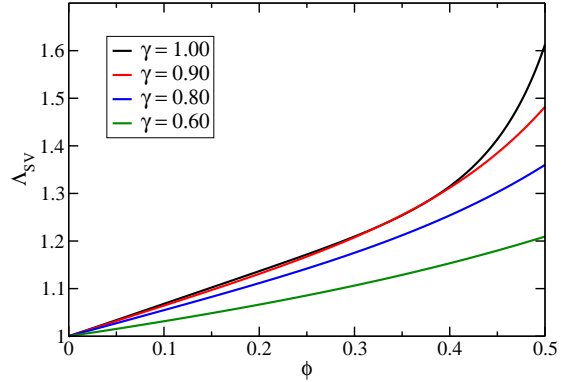}
\caption{Test of the short-time GSE relation in Eq. (\ref{eq:GSEshorttime}) 
for spherical annulus particles with values of $\gamma$ as indicated. Solid lines: 
Product function $\Lambda_\text{SV}$ of the generalized Sait\^o expression 
in Eq. (\ref{eq:Saito_formula}) for $\eta_\infty/\eta_0$ 
and Eq. (\ref{eq:self_diff_improved}) for $D_S/D_0^t$.}
\label{fig:gse_check}
\end{figure}

As straightforward applications of the generalized Sait\^o 
formula and Eq. (\ref{eq:self_diff_improved}) describing $\eta_\infty(\phi,\gamma)$ and  
$D_S(\phi,\lambda)$ in the spherical annulus model, we analyze 
next the validity 
of the short-time generalized Stokes-Einstein (GSE) relation,

\begin{equation} \label{eq:GSEshorttime}
\Lambda_\text{SV}(\phi,\gamma)\approx 1 \,,
\end{equation}  
with the short-time GSE function 
\begin{equation} \label{eq:GSEshorttimefunction}
\Lambda_\text{SV}(\phi,\gamma)=\frac{D_S(\phi,\gamma)}{D_0^\text{t}(\gamma)}
\times\frac{\eta_\infty(\phi,\gamma)}{\eta_0}\,.
\end{equation}  
Eq. (\ref{eq:GSEshorttime}) expresses that $D_S(\phi,\gamma)$ 
should be proportional for all concentrations 
to the inverse of $\eta_\infty(\phi,\gamma)$. 
This relation is trivially fulfilled at infinite dilution where 
it reduces to the single-sphere translational Stokes-Einstein relation 
for a hydrodynamically structured colloidal sphere.   
The approximate validity of this relation would be quite useful, since $\eta_\infty$ can then be determined    
more easily, and using a smaller amount of particles, by a dynamic scattering experiment instead of a rheo-mechanical 
measurement. 
This is why GSE relations including the present one have been thoroughly subjected to theoretical explorations, for particulate systems including permeable hard spheres \cite{Abade2010a} and charge-stabilized particles\cite{Banchio1999,Banchio2008}. 

An exact GSE relation is reflected in Fig. (\ref{fig:gse_check}) by a horizontal straight line of unit height. However, for all considered values of $\gamma$, 
significant deviations from $\Lambda_\text{SV}=1$ are observed at larger volume fractions. 
The deviations are largest for no-slip hard-core particles where the HIs are strongest, 
amounting to about $40\;\!\%$ at $\phi=0.45$. For concentrations $\phi\leq 0.4$, 
the displayed curves for $\Lambda_\text{SV}(\phi)$ are nearly straight lines, 
characterized by the linear coefficient, 
$\lambda_\text{SV}(\gamma)$, 
in the expansion of $\Lambda_{SV}$ to linear order in $\phi$. 
The linear concentration coefficient derived from our analytic expressions for $D_S$ and $\eta_\infty$ 
is given by the polynomial
\begin{equation} \label{eq:lambdasv}
 \lambda_\text{SV}(\gamma) = 0.6685 + 0.3201 \overline{\gamma} + {\cal O}\left(\overline{\gamma}^2\right) \,.  
\end{equation}

For a given hydrodynamic particle model, the values of  
$\gamma$ which should be used in calculating $D_S$ and $\eta_\infty$, respectively, 
are actually different if the 
${\cal O}(L_{h,f}^\ast)^2$ corrections to $a_{h,f}$ in Eq. (\ref{eq:ah-expand}) cannot be neglected.  
However, this does not affect our general conclusion that the GSE relation is violated, 
as illustrated by the curves in Fig. (\ref{fig:gse_check}) where for simplicity 
the same $\gamma$ values were used in calculating $D_S$ and $\eta_\infty$.

According to Fig. (\ref{fig:gse_check}), the ordering relation $\Lambda_\text{SV}(\phi) > 1$ is obeyed  
by particles with pure hard-core interactions. As shown in \cite{Banchio2008}, 
the same ordering applies to 
charge-stabilized suspensions. In contrast, the 
long-time product function $\Lambda_\text{LV}(\phi)=(D_L/D_0^t)\times(\eta/\eta_0)$ relating  
long-time self-diffusion coefficient $D_L$ to zero-frequency viscosity $\eta$  
has been shown for no-slip hard-sphere and charge-stabilized 
suspensions to fulfill the opposite ordering $\Lambda_\text{LV}(\phi) < 1$ for $\phi>0$ \cite{Banchio1999}.

\section{Theory versus Experiment}
\label{sec:theory-experiment}

We demonstrate now the accuracy of our easy-to-apply toolbox methods of calculating short-time 
diffusion properties by 
analyzing DLS measurements by Eckert and 
Richtering \cite{Eckert2008} on non-ionic PNiPAM microgels in DMF. 
As discussed in Subsec. \ref{subec:pair_correlations}, 
the microgel particles behave as hard spheres as far as their static properties are concerned. On modeling 
the microgels also hydrodynamically as no-slip hard spheres with $a_h=a$, and on basis of bare $\delta\gamma$ method results for $H(q)$, Eckert and Richtering came to the conclusion that short-time particle diffusion is underestimated by the no-slip hard-sphere model. This suggests that the non-uniform crosslinker-density of the microgels 
should have a significant hydrodynamic effect.

To account for this effect, we model here the microgels as spherical annulus particles,   
and determine $H(q)$ using the scaling Eq. (\ref{eq:red_hyd_func}) in conjunction with the analytic expressions in  Eqs. (\ref{eq:K_fit}) and (\ref{eq:self_diff_improved}) for $K$ and $D_S$, respectively. 
In the $\delta\gamma$ method calculation of $h_d^{ns}(q)$ which enters 
into the scaling expression of $H(q)$, we use the HS-PY structure factors depicted in Fig. \ref{fig:1_struct}, with the hard-core radius $a=120$ nm. The only adjustable parameter in our model is thus 
the reduced hydrodynamic radius $\gamma=a_h/a$. 
\begin{figure}[hbtp]
\centering
\includegraphics[width=0.5\textwidth]{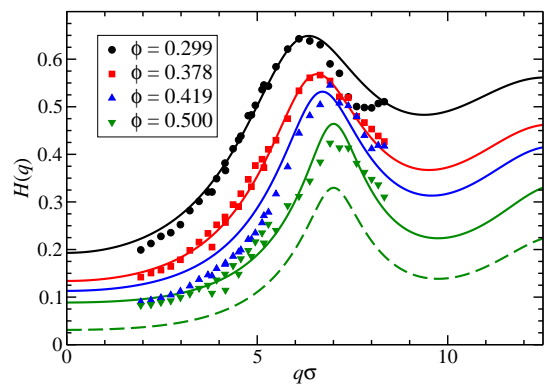}
\caption{ Experimentally deduced hydrodynamic function of PNiPAM 
microgels in DMF (filled symbols, taken from \cite{Eckert2008}) compared with the theoretical 
predictions (solid lines) for the spherical annulus model, using $\gamma=0.97$ 
and Eq. (\ref{eq:red_hyd_func}) for $H(q)$ combined with Eq. (\ref{eq:K_fit}) 
for $K$ and Eq. (\ref{eq:self_diff_improved}) for $D_S$. 
Dashed line: Theoretical prediction for non-permeable particles ($\gamma=1$) at $\phi=0.5$. 
The wavenumbers are scaled by the 
hard-core diamter $\sigma=240$ nm.}
\label{fig:13_hyd_poster_scaling_fit}
\end{figure}

In Fig. \ref{fig:13_hyd_poster_scaling_fit}, our theoretical results for $H(q)$ are presented and compared  
with the experimental findings in \cite{Eckert2008}. The latter have been obtained indirectly from dividing the DLS first cumulant data for the 
diffusion function $D(q)$ by the VW-PY $S(q)$ taken at $q=0$ (c.f. Eq. (\ref{eq:diff_func})). 
Using the constant ratio $\gamma=0.97$, we obtain very good agreement between theory and experiment for all 
volume fractions. Our finding of a concentration-independent hydrodynamic radius $a_h = 0.97\times a$ points to the consistency of our analytic  
method of calculating $H(q)$, since as an
intrinsic particle property, $a_h$ should not depend significantly on the volume fraction. While this holds for the strongly cross-linked non-ionic microgels considered here, for weakly cross-linked ionic microgels 
in the swollen-state temperature range a significant size shrinkage with increasing concentration 
is observed \cite{Holmqvist2012}.        

The deduced hydrodynamic microgel radius is only $3\%$ smaller than the excluded volume radius,  
corresponding to a likewise small value, $\lambda_x = 0.029$, of the reduced fluid penetration length. 
This exemplifies the common 
experimental situation of $a_h \approx a_{h,f}$, 
with a relative correction to the flat plane value $a_{h,f}$ being here 
of ${\cal O}\left((L_{h,f}^\ast)^2\right)\approx 10^{-3}$ small. 

The small microgel permeability nonetheless significantly affects $H(q)$, in particular at larger concentrations. 
This is shown in Fig. \ref{fig:13_hyd_poster_scaling_fit} for $\phi=0.5$ by the comparison with the 
hydrodynamic function for zero permeability (dashed curve): The residual particle permeability enlarges the sedimentation velocity by more than $100 \;\!\%$, and the self-diffusion coefficient by more than $30\;\!\%$.     
\begin{figure}[hbtp]
\centering
\includegraphics[width=0.5\textwidth]{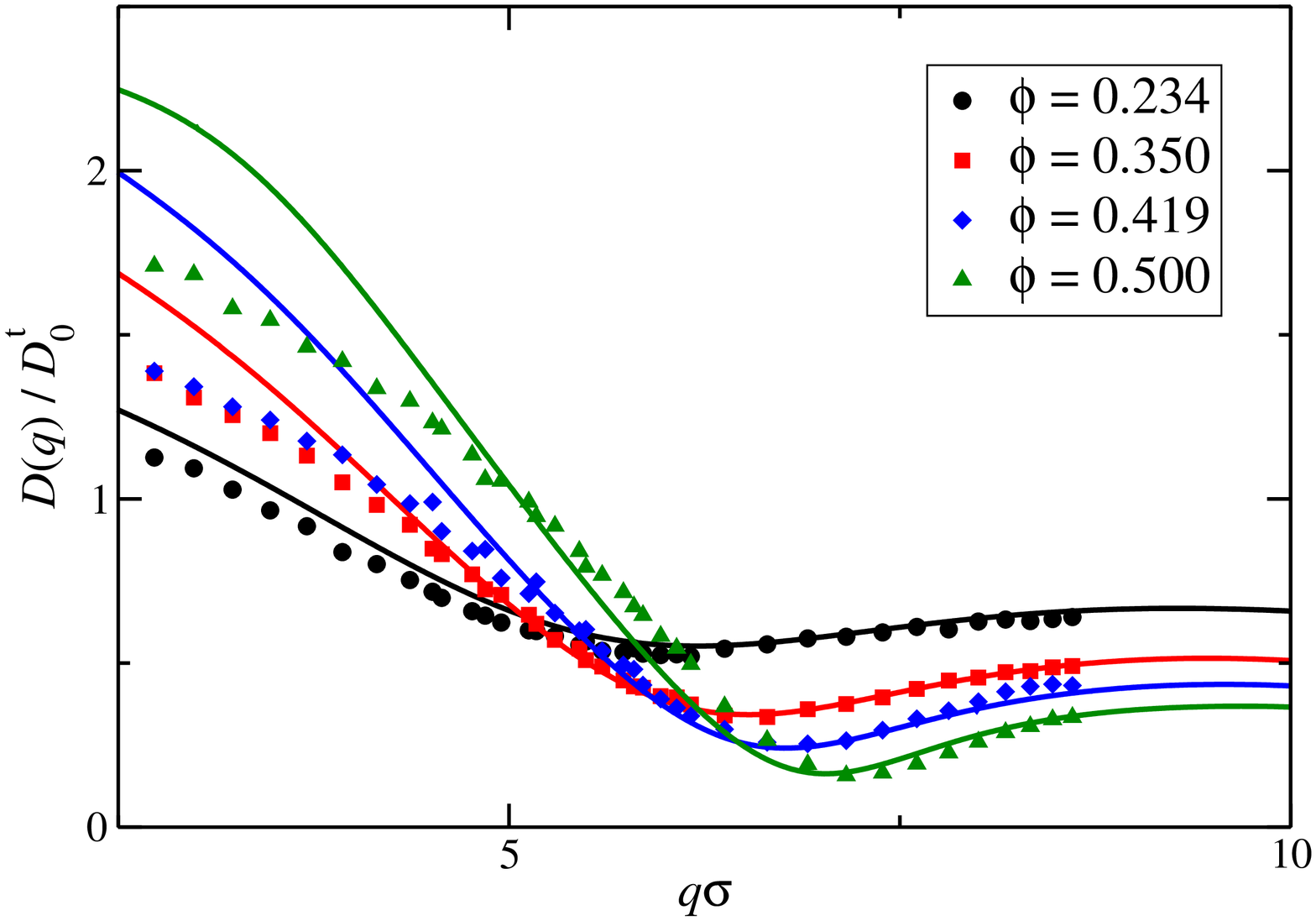}
\caption{DLS data for the reduced diffusion function $D(q)/D_0^t$ (filled symbols) taken from \cite{Eckert2008}, in comparison with our theoretical predictions for $\gamma=0.97$, obtained from the multiplying the sperical annulus results for $H(q)$ depicted in Fig. \ref{fig:13_hyd_poster_scaling_fit} by the VW-PY values of $S(q=0)$.}
\label{fig:14_dif_func}
\end{figure}

In Fig. \ref{fig:14_dif_func}, our theoretical results for $D(q)/D_0^\text{t}$ are plotted together with the DLS data  \cite{Eckert2008} for the same quantity.  
The theoretical curves have been obtained from dividing the spherical annulus $H(q)$'s  
depicted in Fig. \ref{fig:13_hyd_poster_scaling_fit} by the associated VW-PY 
structure factors, 
$S(q)$, of hard spheres in accordance with Eq. (\ref{eq:diff_func}). At $q=0$ , the VW-PY $S(q)$ reduces to the Carnahan-Starling expression for the reduced osmotic compressibility,
\begin{equation}
 S(q=0) = \frac{\left(1-\phi\right)^4}{\left(1 +2\;\!\phi\right)^2 +\phi^3 \left(\phi-4\right)} \,,
\end{equation}  
valid in the full fluid-phase volume fraction range of hard spheres.  
The agreement between theoretical and experimental diffusion functions is very good in the intermediate wavenumber range including the principal peak position, $q_m$, of $S(q)$ where $D(q)$ is minimal, and also for larger wavenumbers. 
At small $q$ values and large volume fractions, the experimental data are overestimated. 
Even considering the remaining small-$q$ deviations, the here presented theoretical 
results for $D(q)$ are in distinctly better agreement with the experimental data than the 
earlier ones presented in \cite{Eckert2008} where permeability effects were not included.  

The deviations in $D(q)$ at small-$q$ can be partially attributed to the high sensitivity of the inverse  
compressibility factor, $1/S(0)$, multiplying $H(0)$ in Eq. (\ref{eq:dcoll}), 
on the residual softness of the microgels.  
\begin{figure}[hbtp]
\centering
\includegraphics[width=0.5\textwidth]{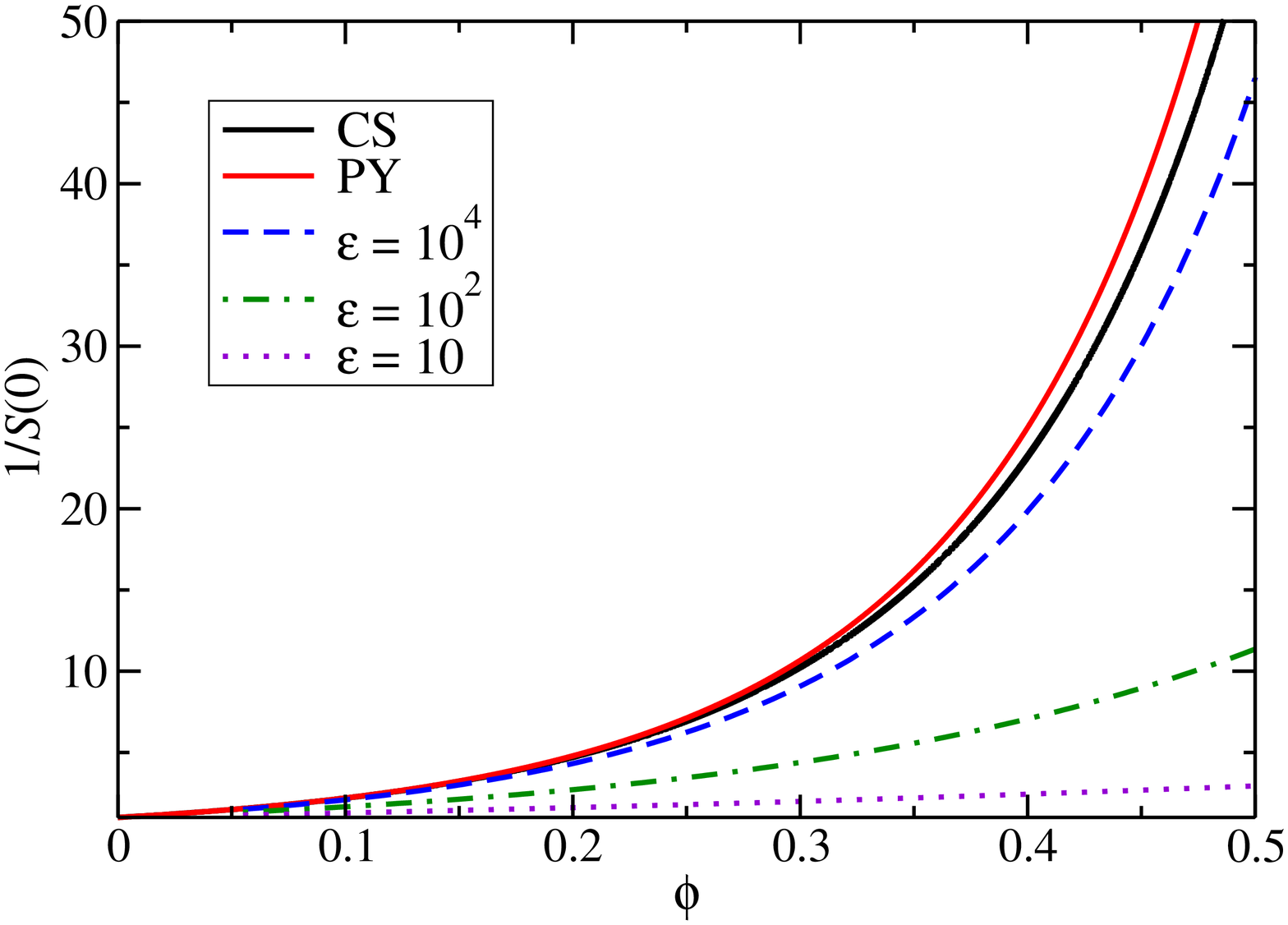}
\caption{Influence of particle softness (elasticity) on the inverse compressibility factor $1/S(0)$, 
for hard-sphere and Hertz potential particles, plotted as a function of $\phi$. The Carnahan-Starling (CS) (black solid line) and PY (red solid line) results for hard spheres are compared with the PY-based pedictions for Hertz model particles with softness parameters $\epsilon=10^4$
(dashed line), $10^2$ (dashed-dotted line), and $10$ (dotted line).}
\label{fig:15_S_in_zero_inverse}
\end{figure}
This is demonstrated in Fig.  
\ref{fig:15_S_in_zero_inverse}, where the concentration dependence of $1/S(0)$  
for hard spheres is compared to that of the Hertz potential system for the strongly distinct softness parameters 
$\epsilon=10^{4}$, $10^2$ and $10$. According to the figure, even a small residual softness characterized by $\epsilon=10^{4}$ significantly enlarges the osmotic compressibility for large volume fractions, 
with $1/S(0)$ being lowered accordingly. Viewed on the scale of the structure factors in Fig. \ref{fig:2_S_378}, 
the small-$q$ differences are not resolved since $S(0)$ is very small for large concentrations. While the smaller factor $1/S(0)$ in the Hertz model would improve the agreement with the experimental $D(q)$ at small $q$, 
we recall in referring to Fig. \ref{fig:2_S_378} that a somewhat larger volume fraction $\phi_s$ than for hard spheres is 
required in the Hertz model for an equally good fit of the experimental $S(q)$. If the enlarged volume fraction is accounted for, there remains a small final reduction in $1/S(0)$ only. An additional cause for the small-$q$ differences can be size polydispersity. It was shown in \cite{Phalakornkul1996},
that a small degree of polidispersity significantly enlarges in concentrated suspensions  
the measured diffusion function at small $q$.   

In summarizing, we conclude that our HRM based toolbox methods   
reproduce the short-time diffusion properties of the non-ionic microgel suspensions very well, 
and with little numerical effort.  
The non-homogeneous cross-linker density 
is accounted for by a hydrodynamic radius which is only three percent 
smaller than the excluded volume radius.

\section{Long-time Transport Properties}
\label{sec:long-time-properties}

Long-time colloidal transport properties such as $D_L$ and $\eta$ characterize the particle dynamics 
on time scales $t \gg \tau_D$ during which 
the particle configuration has changed significantly. Different from short-time properties, 
they are influenced by thermally driven microstructural relaxations 
depending on direct and hydrodynamic interactions alike. This renders a first-principles calculation of long-time properties 
demanding, both in theory and simulations, in particular when the salient HIs are accounted for.  
Consequently, in most simulation studies of the concentration dependence of $D_L$ and $\eta$, 
the influence of the HIs  has been ignored \cite{Cichocki1992}\cite{Strating1999,Foss2000}. 
The few existing three-dimensional 
simulation studies where HIs are included have been focused on Brownian hard spheres 
with no-slip BC \cite{Phung1993,Foss2000a,Banchio2003}.
Therefore, a theoretical scheme is on demand allowing for the approximate calculation of $D_L$ and $\eta$ 
for suspensions of hydrodynamically structured particles.   

In the following, we present such a scheme which as a bonus requires only little numerical effort. 
It is based on the HRM and a 
factorization approximation method proposed originally by Medina-Noyola for self-diffusion \cite{Medina-Noyola1988}. 
We point out that the HRM error estimation for short-time properties in Eq. (\ref{eq:HRM_Hofq_estimate}) applies also to long-time  properties including $D_L/D_0^\text{t}$ and $\eta$, and also the dynamic structure factor $S(q,t)$ for arbitrary correlation times. This follows from general expressions 
for long-time transport coefficients and $S(q,t)$ which have been obtained using the Mori-Zwanzig projection operator formalism in conjunction with the many-particle generalized Smoluchowski equation for the configurational probability distribution function \cite{Nagele1996}. The crucial fact to notice here is that the hydrodynamic mobility tensors entering into the GSmE have no explicit time dependence. 

We exemplify the error estimate for $D_L$ by starting 
from the configurational average expression \cite{Nagele1996},
\begin{eqnarray} \label{eq:Dlongtheory}
D_L  =  D_S+\Delta D \,,
\end{eqnarray}        
with 
\begin{eqnarray} \label{eq:DeltaDlong}
\Delta D = -\left\langle (\hat{\bf q}\cdot{\bf V}_1) {\cal O}_B^{-1} (\hat{\bf q}\cdot{\bf V}_1) \right\rangle \,.
\end{eqnarray}        
The slowing effect on $D_L$ by the dynamically restructuring cage of next neighbors formed around each particle is embodied in the negative valued relaxation contribution, $\Delta D$, to $D_L$ implying $D_L < D_S$ for $\phi>0$.       
In Eq. (\ref{eq:DeltaDlong}), ${\bf V}_1 = {\cal O}_B {\bf R}_1$ is the coarse-grained velocity of a 
representative particle $1$ whose Brownian motion is considered, $\hat{\bf q} = {\bf q}/q$, and ${\cal O}_B({\bf X})$ the so-called backward Smoluchowksi differential operator generating the time evolution of the probability distribution function. 
The operator inverse is denoted by ${\cal O}_B^{-1}$. The only information about ${\cal O}_B$ needed here is its linear dependence on the mobility tensors $\bm{\mu}_{ij}$. From  this and Eq. (\ref{eq:muexpansion}), it follows that the error introduced on approximating $D_L/D_0^\text{t}$ by $D_{L;\text{HRM}}/D_0^\text{t}$ is of ${\cal O}\left((L_{h,f}^\ast)^2\right)$. 

\subsection{Long-time self-diffusion}
\label{subsec:self-diffusion-longtime}

Using Eq. (\ref{eq:Dlongtheory}), $D_L$ can be written as  
\begin{equation}
D_\text{L}\left(\phi\right) = D_{S}\left(\phi\right)
\left[ 1+\frac{\Delta D(\phi)}{D_S(\phi)}\right]\,,
\label{eq:D_long_scaling}
\end{equation}
where in the term in brackets, the explicit dependence on $D_S$ has been scaled out. 
According to arguments first put forward by Medina-Noyola \cite{Medina-Noyola1988}, and 
subsequently substantially elaborated by Brady also regarding the zero-frequency viscosity \cite{Brady1993,Brady1994}, the factor function in brackets 
is not only scale invariant with respect to $D_S$, but for hard spheres it is to decent approximation 
also independent of the HIs. This implies the no-HI factorization approximation, 
\begin{equation}
\left[ 1+\frac{\Delta D}{D_S}\right] \approx \left[ \frac{D_L}{D_0^\text{t}} \right]_\text{no-HI}\,,
\label{eq:D_long_scaling_2}
\end{equation}
where the bracket term is approximated by a purely structural factor determined by excluded volume interactions only. For known $D_S$, the problem of calculating $D_L$ is thus simplified to the calculation of the reduced long-time self-diffusion coefficient without HIs. The dependence of $D_L$ on the HIs, 
and the hydrodynamic particle structure, is embodied here in $D_S$ alone. 

\begin{figure}[hbtp]
\centering
\includegraphics[width=0.5\textwidth]{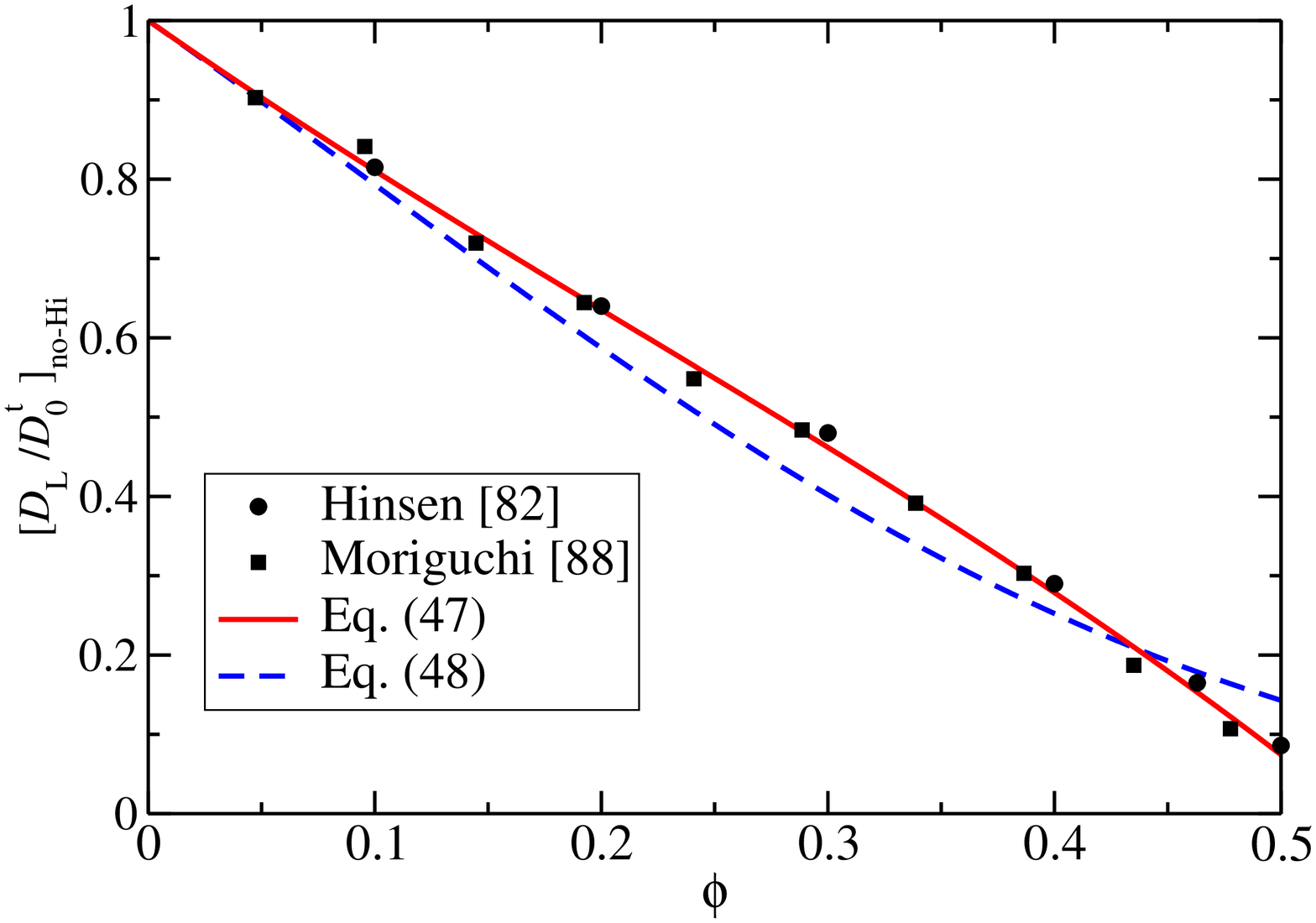}
\caption{Reduced long-time self-diffusion coefficient, $\left[ D_L/D_0^\text{t} \right]_\text{no-HI}$, 
of colloidal hard spheres without HIs. Filled circles and squares: 
Brownian dynamics simulation results by Hinsen and Cichocki \cite{Cichocki1992} and Moriguchi \cite{Moriguchi1997}, respectively. 
Solid line: Polynomial fit in Eq. (\ref{eq:Dlong-fit-noHI}). 
Dashed line: Eq. (\ref{eq:Brady-DL-noHI}) 
with Carnahan-Starling contact value input.}
\label{fig:D_long_no-HI}
\end{figure}

Brownian dynamics simulation results for $[D_L(\phi)/D_0^\text{t}]_\text{no-HI}$ by Hinsen and Cichocki \cite{Cichocki1992} and Moriguchi \cite{Moriguchi1997} are depicted in Fig. \ref{fig:D_long_no-HI}. In the fluid-phase concentration regime $\phi \leq \phi_f$, where $\phi_f = 0.494$ is the volume fraction at freezing, the simulation data are well described by the polynomial least-square fit, 
\begin{equation} \label{eq:Dlong-fit-noHI}
  \left[\frac{D_L}{D_0^\text{t}} \right]_\text{no-HI} = 1-2\;\!\phi+1.272\;\!\phi^2-1.951\;\!\phi^3 
 \end{equation}
where the exact first-order virial coefficient, $\lambda_L = 2$, for a hard-sphere suspension without HIs has been incorporated.  
The figure depicts furthermore the analytic approximation, 
\begin{equation} \label{eq:Brady-DL-noHI}
  \left[ \frac{D_L}{D_0^\text{t}} \right]_\text{no-HI} \approx \frac{1}{1 + 2\:\!\phi g(\sigma^+;\phi)} \,,
\end{equation}
given by Brady \cite{Brady1993}, where the structural factor is expressed in terms of the RDF contact value, 
$g(\sigma^+;\phi)$, of hard spheres. The contact value for the fluid-phase concentration range 
is to high accuracy described by the Carnahan-Starling expression \cite{Rintoul1996} 
\begin{equation} \label{eq:gcontCS}
  g(\sigma^+;\phi) = \frac{1-\phi/2}{\left(1 - \phi\right)^3} \,
\end{equation}
with $g(\sigma^+;\phi_f)=5.81$. 
Eq. (\ref{eq:Brady-DL-noHI}) incorporates the exact first-order virial coefficient, $\lambda_L= -2$.  
Moreover, near random closed packing at   
$\phi_\text{rcp} \approx 0.64$ where a metastable hard-sphere fluid gets jammed, 
on basis of results for $g(\sigma^+;\phi)$ by Rintoul and Torquato \cite{Rintoul1996} 
it predicts that    
$\left[D_L\right]_\text{no-HI}$ diminishes linearly    
like $0.59\times(1 - \phi/\phi_\text{rcp})$. Since $D_S$     
vanishes likewise linearly in case of no-slip hard spheres, 
the quadratic scaling prediction $D_L \sim (1 - \phi/\phi_\text{rcp})^2$ near random closed packing 
is obtained in the factorization approximation. 

\begin{figure}[hbtp]
\centering
\includegraphics[width=0.5\textwidth]{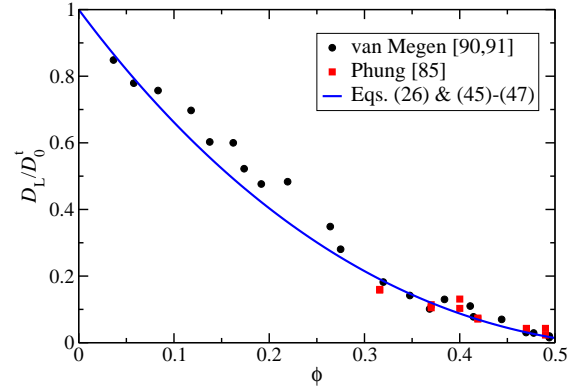}
\caption{Reduced long-time self-diffusion coefficient, $D_L/D_0^\text{t}$, of no-slip colloidal hard spheres. Filled circles:  Experimental data by van Megen {\em et al.} \cite{VanMegen1986,VanMegen1989}. Filled squares: Simulation data 
by Phung {\em et al.} \cite{Phung1993}. Solid line: No-HI factorization approximation 
in Eqs. (\ref{eq:D_long_scaling})-(\ref{eq:Dlong-fit-noHI}), 
with the short-time factor $D_S/D_0^t$  
according to Eq. (\ref{eq:self_diff_improved}) for $\gamma=1$.}
\label{fig:16_D_long_HI_comp}
\end{figure}

The performance of the no-HI factorization approximation for the $D_L$ of no-slip spheres 
on basis of Eq. (\ref{eq:self_diff_improved}) for $D_S$ at $\gamma=1$ and Eq. (\ref{eq:Dlong-fit-noHI})  
for $\left[ D_L/D_0^t \right]_\text{no-HI}$, is documented in Fig. \ref{fig:16_D_long_HI_comp} 
by the comparison with DLS data by van Megen {\em et al.} \cite{VanMegen1986,VanMegen1989},  
and simulation results by Phung {\em et al.} \cite{Phung1993} with HIs included. 
The Stokesian dynamics simulation data for Brownian hard spheres by Phung {\em et al.} 
have been obtained for a small number ($N=27$) of particles, and for a small 
albeit non-zero shear Peclet number $\text{Pe} = 0.01$.    
The overall agreement with the experimental and simulation data is quite good.   
The factorization approximation gives $\lambda_L = -3.831$ for the first-order virial coefficient 
of hydrodynamically interacting no-slip spheres, while its correct numerical value is given by $\lambda_L = -2.1$. 
The initial low-concentration decrease of $D_L$ is thus overestimated. 

\begin{figure}[hbtp]
\centering
\includegraphics[width=0.5\textwidth]{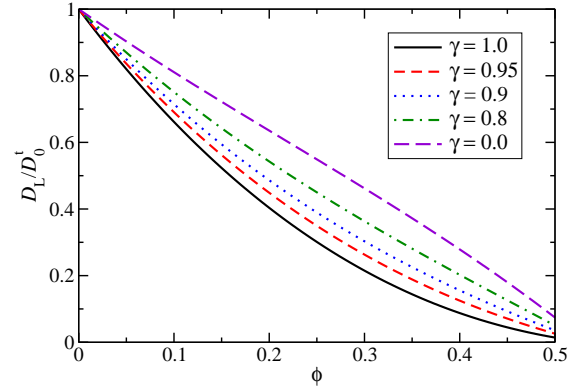}
\caption{Generic influence of the intra-particle hydrodynamic structure on $D_L/D_0^t$,  
quantified in the no-HI factorization approximation, Eqs. (\ref{eq:D_long_scaling}) and (\ref{eq:Dlong-fit-noHI}), 
using for $D_S/D_0^t$ the spherical annulus model formula in 
Eqs. (\ref{eq:self_diff_improved}) and (\ref{eq:sa_virial_fit}). Several values of $\gamma$ are 
considered as indicated.}
\label{fig:17_D_long_HI_perm}
\end{figure}

The good performance of the no-HI factorization approximation for no-slip hard spheres 
gives support to its straightforward extension to hydrodynamically structured particles, 
by using for $D_S/D_0^t$ now the analytic expression in Eq. (\ref{eq:self_diff_improved}) for spherical annulus spheres. Our results for $D_L/D_0^\text{t}$ based on this extended factorization scheme are shown in 
Fig. \ref{fig:17_D_long_HI_perm}. With decreasing 
$\gamma$, the slowing down effect of the HIs on $D_L$ diminishes. In the limit $\gamma\to 0$ of hard spheres acting hydrodynamically as point particles, the long-time self-diffusion coefficient in the absence of HIs is recovered. 
According to Cichocki and Felderhof \cite{Cichocki1991}, the linear concentration coefficient 
of $D_L/D_0^\text{t}$ for small $\gamma$ is $\lambda_L(\gamma)= -2\left[1 - 1.031\;\!\gamma +0.111\;\!\gamma^2 
+ {\cal O}(\gamma^3) \right]$, where the linear and quadratic terms in $\gamma$ are due to the so-called 
Oseen long-distance HIs contribution to the relaxation part $\Delta D$. 
The contributions of the short-time part, $D_S$, to $D_L$ appear first in quadratic order in $\gamma$. 
While this describes quantitatively how $D_L$ approaches $[D_L]_\text{no-HI}$ for small $\gamma$, 
we reemphasize that 
$\gamma > 0.8$ for most hydrodynamically structured colloidal particles.   

The no-HI factorization approximation predicts the ratio, $D_L/D_S$, 
of long-time and short-time coefficients to be independent 
of HIs and hydrodynamic particle structure, with value equal to $[D_L/D_0^t]_\text{no-HI}$. 
This implies with Eqs. (\ref{eq:D_long_scaling}) and (\ref{eq:D_long_scaling_2}) that 
\begin{equation}
 \left(\frac{D_L}{D_S}\right)(\phi_f) \approx 0.1 \,,
\end{equation} 
in good accord with the L\"owen-Palberg freezing criterion value of about $0.1$. Thus, 
a universal freezing value of $0.1$ is predicted not only 
for colloidal suspensions with different pair potentials, but also with different hydrodynamic intra-particle structures. 
For pair interactions characterized by a single length scale, the dynamic L\"owen-Palberg criterion has been shown 
to be equivalent to the static 
Hansen-Verlet freezing criterion for the value $S(q_m)$ of the structure factor peak height.

\subsection{Zero-frequency viscosity}
\label{subsec:steady-shear-viscosity}

Analogous to Eq. (\ref{eq:D_long_scaling}) for $D_L$, in Eq. (\ref{eq:zero_freq}) for the low-shear zero-frequency viscosity $\eta$, we factor out the high-frequency (short-time) contribution $\eta_\infty$ according to 
\begin{equation}
\eta\left(\phi\right) = \eta_\infty\left(\phi\right)\left[1+\frac{\Delta\eta\left(\phi\right)}{\eta_\infty\left(\phi\right)}\right]\,,
\label{eq:zero_freq_scal}
\end{equation}
with the term in brackets expected to be approximately independent of the HIs. 
This suggests the no-HI factorization approximation, 
\begin{equation}
\frac{\Delta\eta\left(\phi\right)}{\eta_\infty\left(\phi\right)} 
\approx \left[\frac{\Delta\eta\left(\phi\right)}{\eta_0}\right]_\text{no-HI}\,,
\label{eq:visc-fac-approx}
\end{equation}
where $[\Delta\eta]_\text{no-HI}$ is the shear relaxation viscosity part without HIs. 
In this approximation, 
the HIs are assumed to affect $\eta$ only by means of the factored out $\eta_\infty$ in Eq. (\ref{eq:zero_freq_scal}). 
The neglect of HIs simplifies the calculation of the shear relaxation viscosity part considerably.  
Following works by Brady \cite{Brady1993,Brady1994}, an analytic estimate of $\Delta\eta$   
for no-slip hard spheres without HIs is given by
\begin{equation} \label{eq:Deltaeta-noHI}
 \left[\frac{\Delta\eta}{\eta_0}\right]_\text{no-HI} \approx \frac{12}{5}\phi^2 g(\sigma^+;\phi)\,.
\end{equation}
\begin{figure}[hbtp]
\centering
\includegraphics[width=0.5\textwidth]{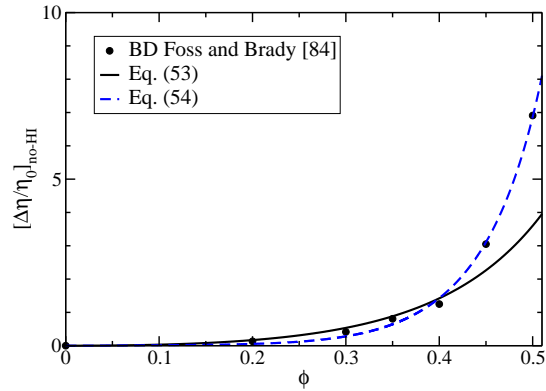}
\caption{Reduced shear relaxation viscosity part without HIs, $[\Delta\eta/\eta_0]_\text{no-HI}$, 
of a suspension of Brownian hard spheres in dependence of $\phi$. Filled circles: Brownian dynamics (BD) simulation data by Foss and Brady \cite{Foss2000}. Solid line: Analytic estimate in Eq. (\ref{eq:Deltaeta-noHI}). 
Dashed line: Semi-empirical fit in Eq. (\ref{eq:Deltaeta-noHI-fit}).} 
\label{fig:zeroviscosity_no_HI}
\end{figure}
This estimate combines the exact low concentration limit, $2.4\:\!\phi^2 +{\cal O}(\phi^3)$, of $[\Delta\eta]_\text{no-HI}$ 
with its divergence at random closed packing according to $[\Delta\eta]_\text{no-HI}\sim\left(1-\phi/\phi_\text{rcp}\right)^{-1}$, triggered by the divergence of the hard-sphere contact value. 
Together with the likewise linear divergence of $\eta_\infty$ for no-slip hard spheres, 
a quadratic divergence $\eta\sim\left(1-\phi/\phi_\text{rcp}\right)^{-2}$ is thus 
predicted for the zero-frequency viscosity. For hydrodynamically structured particles 
where $a_h < a$, there are no diverging lubrication forces for spheres in contact. 
The high-frequency viscosity remains then finite at random closed packing, 
and the particles can still rotate individually.  
Moreover, $\eta$ diverges only 
linearly as $\eta(\phi,\gamma<1)\sim\left(1-\phi/\phi_\text{rcp}\right)^{-1}$.

In Fig. \ref{fig:zeroviscosity_no_HI}, the outcome of Eq. (\ref{eq:Deltaeta-noHI}) 
for $[\Delta\eta]_\text{no-HI}$ is compared to Brownian dynamics simulation results without HIs by Foss and Brady \cite{Foss2000,Foss2000a}. Up to $\phi \approx 0.35$, the simulation data are 
decently well represented by the analytic expression, but the steep rise of $[\Delta\eta]_\text{no-HI}$ 
at large volume fractions is not reproduced.  
The simulation data are well captured for all $\phi$ by the semi-empirical expression 
\begin{equation} \label{eq:Deltaeta-noHI-fit}
\left[\frac{\Delta\eta}{\eta_0}\right]_\text{no-HI} = \frac{\frac{12}{5}\phi^2\left(1-7.085\phi+20.182\phi^2\right)}{\left(1-\frac{\phi}{\phi_\text{rcp}}\right)}\,,
\end{equation} 
combining the exact quadratic-order concentration dependence with the linear order divergence at random closed packing. 
Throughout this work, 
we restrict our analysis to the equilibrium fluid-phase concentration regime $\phi \leq 0.5$, 
while the viscosity simulations in 
\cite{Foss2000,Foss2000a,Strating1999} have been extended to the metastable fluid 
concentration regime $\phi_\text{f} < \phi < \phi_\text{rcp}$ 
where crystallization is kinetically suppressed.      
\begin{figure}[hbtp]
\centering
\includegraphics[width=0.5\textwidth]{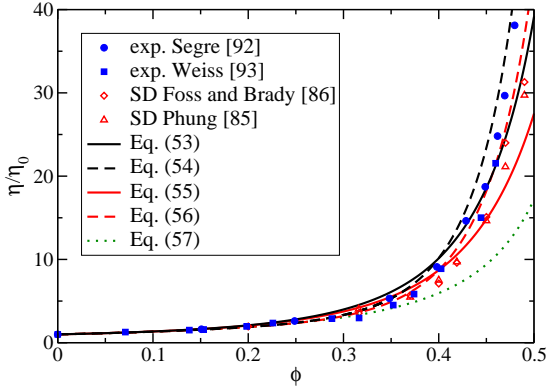}
\caption{Zero-frequency viscosity, $\eta/\eta_0$, of no-slip Brownian hard spheres with HIs. 
Solid and dashed black lines: No-HI factorization predictions using $\eta_\infty(\phi,\gamma=1)$ 
according to Eq. (\ref{eq:Saito_formula}), and $[\eta/\eta_0]_\text{no-HI}$ 
according to Eqs. (\ref{eq:Deltaeta-noHI}) and (\ref{eq:Deltaeta-noHI-fit}), respectively. 
Solid and dashed red lines: Brady's (modified) scaling approximations using 
Eqs. (\ref{eq:Bradyvisc}) and (\ref{eq:Bradyvisc-mod}) for $\Delta\eta/\eta_\infty$, respectively, 
and Eq. (\ref{eq:Saito_formula}) for the factored out $\eta_\infty$. 
The factor $1/\Lambda_\text{SV}$ in the (modified) scaling approximation is calculated using Eq. (\ref{eq:Saito_formula}) for $\eta_\infty$ 
and Eq. (\ref{eq:self_diff_improved}) for $D_S$. Dotted line: Eq. (\ref{eq:viscNaegele}). Filled symbols: Experimental data by Segr\`e {\em et al.} and Weiss {\em et al.} \cite{Segre1995a,Weiss1998}. 
Open symbols: Stokesian dynamics (SD) simulation results for Brownian hard spheres by Foss and Brady \cite{Foss2000a}
and Phung \cite{Phung1993}.}
\label{fig:19_vis_full_brady}
\end{figure}
Regarding the shear relaxation viscosity part with HIs included, Brady has proposed the 
following approximate scaling expression \cite{Brady1993,Brady1994}, 
\begin{equation} \label{eq:Bradyvisc}
 \frac{\Delta\eta}{\;\eta_\infty} 
 \approx \frac{12}{5}\;\!\phi^2\;\!\frac{g(\sigma^+;\phi)}{\Lambda_\text{SV}(\phi)}\,,
\end{equation}
where the influence of the HIs on the ratio $\Delta\eta/\eta_\infty$ is solely embodied in the 
short-time GSE function $\Lambda_\text{SV}$ defined in Eq. (\ref{eq:GSEshorttimefunction}). 
For no-slip hard spheres,  
$\Lambda_\text{SV}$ is well represented for $\phi < 0.4$ by   
$\Lambda_\text{SV} \approx 1 +0.67 \phi$ according to Eq. (\ref{eq:lambdasv}),  
while without HIs $\Lambda_\text{SV}$ is equal to one. 
Consequently, Eq. (\ref{eq:Bradyvisc}) predicts $\Delta\eta/\eta_\infty$ 
to be only mildly affected by the HIs, giving some credit 
to the no-HI factorization approximation in Eq. (\ref{eq:visc-fac-approx}). 
Eq. (\ref{eq:Bradyvisc}) was obtained from arguing that the adequate diffusion time scale in a concentrated suspension 
is $a_h^2/D_s$ instead of $a_h^2/D_0^\text{t}$, and from using a low-concentration estimate of the weakly shear-distorted stationary pair distribution function with the prefactor $g(\sigma^+;\phi)$ preserved \cite{Brady1993}. A detailed 
discussion of the approximations going into Eq. (\ref{eq:visc-fac-approx}) 
is given by Lionberger and Russel \cite{Lionberger1997,Lionberger1999}.  

On using in place of Eq. (\ref{eq:Deltaeta-noHI}) 
the semi-empirical fitting expression in Eq. (\ref{eq:Deltaeta-noHI-fit}) 
as the non-hydrodynamic factor in Eq. (\ref{eq:Bradyvisc}) , 
Brady's scaling relation is modified to 
\begin{equation} \label{eq:Bradyvisc-mod}
 \frac{\Delta\eta}{\eta_\infty} \approx 
 \frac{\frac{12}{5}\phi^2\left(1-7.085\phi+20.182\phi^2\right)}
 {\left(1-\frac{\phi}{\phi_\text{rcp}}\right) \Lambda_\text{SV}(\phi)} \,.
\end{equation}

The two here considered variants of the no-HI scaling approximation of $\eta$ 
consist of using Eqs. (\ref{eq:Deltaeta-noHI}) and (\ref{eq:Deltaeta-noHI-fit}), respectively, 
as input for the ratio $\Delta\eta/\eta_\infty$ 
in the bracket term in Eq. (\ref{eq:zero_freq_scal}), 
in conjunction with the 
accurate generalized Sait\^o formula in Eq. (\ref{eq:Saito_formula}) used for $\eta_\infty$. 
In addition, Brady's scaling expression for $\eta$ and its modification consist of 
approximating $\Delta\eta/\eta_\infty$ in the bracket term 
in Eq. (\ref{eq:zero_freq_scal}) by Eqs. (\ref{eq:Bradyvisc}) and (\ref{eq:Bradyvisc-mod}), respectively, 
with the generalized Saito formula used again for the factored out 
high-frequency viscosity. The hydrodynamic factor $1/\Lambda_\text{SV}$ in the two scaling expressions 
is calculated analytically using Eq. (\ref{eq:self_diff_improved}) for $D_S$, 
and the generalized Sait\^o formula for $\eta_\infty$.    

The results for $\eta(\phi,\gamma=1)$ by the four inter-related analytic approximations 
are depicted in Fig. \ref{fig:19_vis_full_brady}. 
They are compared with experimental data by Segr\`e {\em et al.} \cite{Segre1995a} 
and Weiss {\em et al.} \cite{Weiss1998}, and Stokesian dynamics (SD) simulation data for Brownian hard spheres 
with HIs included by Foss and Brady \cite{Foss2000a} and Phung {\em at al.} \cite{Phung1993}.
The latter have been   
obtained for a small number ($N=27$) of particles in the basic simulation box. In 
\cite{Foss2000a}, $\Delta\eta$ was deduced using a general Green-Kubo formula for the shear stress 
correlation function of hydrodynamically interacting particles \cite{Nagele1998a}. The viscosity  curves by all four approximations compare overall quite well with the simulation data. 
Depending on the used approximation for $\Delta\eta/\eta_\infty$, deviations are visible 
at intermediate and large volume fractions. The neglect of HIs in $\Delta\eta/\eta_\infty$ results in an overestimation of the simulations data at large $\phi$, but the large-$\phi$ experimental viscosity data are well described. On considering the simulation data to be more trustworthy than the experimental data, owing to experimental polydispersity effects and difficulties in determining the precise volume fraction, the modified scaling expression 
by Brady in Eq. (\ref{eq:Bradyvisc-mod}) provides the overall best description of the SD data for $\eta$, 
slightly better than Brady's original scaling expression in Eq. (\ref{eq:Bradyvisc}). 
We use the modified scaling expression in the following discussion of hydrodynamically structured particles. 
Note that the second-order in concentration coefficient of $\eta$ is predicted by all four factorization approximation variants as $5.01 + 2.4 = 7.41$, 
while the exact coefficient is equal to $5.931$ \cite{Cichocki1990}. 
This low-$\phi$ difference is not resolved on the scale of Fig. \ref{fig:19_vis_full_brady}.

The figure includes additionally the viscosity prediction by the 
formula
\begin{equation} \label{eq:viscNaegele}
 \eta(\phi)/\eta_0 = \frac{1 - 0.4 \overline{\phi}+0.222 \overline{\phi}^2}{\left(1 - \overline{\phi}\right)^2} \,, 
\end{equation}
with $\overline{\phi} = \phi/\phi_\text{rcp}$, which incorporates the first two known virial coefficients 
in $\eta/\eta_0 = 1 + 2.5 \phi + 5.91 \phi^2 +{\cal O}(\phi^3)$ and a quadratic divergence of $\eta$ 
at $\phi_\text{rcp}$. As seen in Fig. \ref{fig:19_vis_full_brady}, 
it describes the simulation and experimental data well for $\phi \leq 0.35$, but it strongly underestimates them at larger 
$\phi$. 

\begin{figure}[hbtp]
\centering
\includegraphics[width=0.5\textwidth]{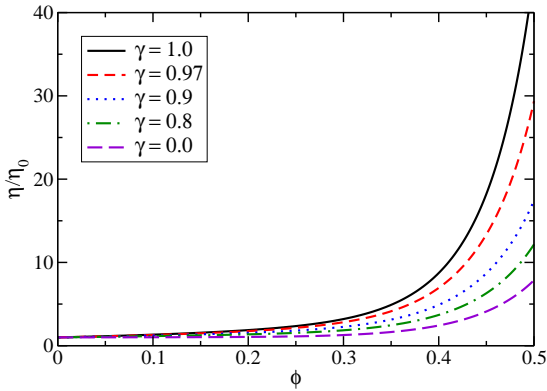}
\caption{Predictions for 
the reduced zero-frequency viscosity, $\eta(\phi,\gamma)/\eta_0$, of hydrodynamically structured particles, based on the modified Brady scaling expression for $\Delta\eta/\eta_\infty$ in Eq. (\ref{eq:Bradyvisc-mod}), and the generalized Sait\^o formula in Eq. (\ref{eq:Saito_formula}) for $\eta_\infty(\phi,\gamma)$. The hydrodynamic factor $\Lambda_\text{SV}(\phi,\gamma)$ is calculated using Eq. (\ref{eq:Saito_formula}) for $\eta_\infty(\phi,\gamma)$, and Eq. (\ref{eq:self_diff_improved}) 
for $D_S(\phi,\gamma)/D_0^t(\gamma)$. Several values of $\gamma$ are considered as indicated.
}
\label{fig:20_vis_full_gamma}
\end{figure}

Akin to long-time self-diffusion, we can straightforwardly extend our analysis to 
hydrodynamically structured particles, by using the generalized Sait\^o expression  
in Eq. (\ref{eq:Saito_formula}) for $\eta_\infty(\phi,\gamma)$, 
in combination with the modified Brady scaling expression in 
Eq. (\ref{eq:Bradyvisc-mod}) for $\Delta\eta/\eta_\infty$. The viscosity predictions 
for different reduced hydodynamic radii $\gamma$ and volume fractions 
extending up to $\phi_\text{f}$ 
are depicted in Fig. \ref{fig:20_vis_full_gamma}. Note the pronounced reduction 
of $\eta(\phi,\gamma)$ with decreasing $\gamma$, owing to the reduced dissipation,  
inherited from the similar behavior of $\eta_\infty(\phi,\gamma)$. In the limit $\gamma \to 0$, 
the zero-frequency viscosity reduces to $\eta(\phi,0) = \eta_0 +[\Delta\eta]_\text{no-HI}$. 
In spite of being approximate, 
the analytic modified scaling expression for $\eta(\phi,\gamma)$ in Eq. (\ref{eq:Bradyvisc-mod}) 
can be expected to be useful for a quick analysis of experimental viscosity data of hydrodynamically structured colloidal suspensions. For the PNiPAM microgels in DMF, e.g., where $\gamma=0.97$ has been deduced, 
a significant reduction both in $\eta$ and $\eta_\infty$ is predicted 
relative to the corresponding viscosities of non-permeable particles. 
It will be interesting to compare our zero-frequency and high-frequency 
viscosity predictions with future viscosity measurements on non-ionic PNiPAM in DMF systems.     

\section{Concluding Remarks}
\label{sec:conclusions}

We have presented a toolbox of methods for calculating short-time and long-time transport properties of suspensions of spherical particles with intrinsic hydrodynamic structure. The analytic scaling expressions given in the paper combine high accuracy predictions with a facile implementation, and they apply to the full liquid-phase concentration regime. They are useful to experimenters for a fast yet precise data analysis of scattering and rheo-mechanical experiments. We have highlighted this in our analysis of SLS and DLS experiments on non-ionic PNiPAM microgels in DMF. By a detailed comparison with Hertz potential calculations, we have shown that the microgels behave statically to good accuracy as hard spheres, and hydrodynamically as permeable spheres with a reduced penetration length $\lambda_x \approx 0.03$  corresponding to a reduced hydrodynamic radius of $\gamma=0.97$. It will be interesting to scrutinize the analytic viscosity expressions for $\eta_\infty$ and $\eta$ of hydrodynamically structured, rigid spheres against future rheo-mechanic  measurements on non-ionic PNiPAM microgels.

The analytic toolbox expressions for the collective diffusion coefficient and the zero-frequency viscosity have been already profitably used in a recent crossflow ultrafilration study of solvent-permeable nanoparticles suspensions \cite{Roa}. The short-time transport coefficient expressions can serve also as input in the calculation of frequency- and time-dependent transport properties on basis of mode-coupling theory and dynamic density functional theory methods were HIs are included \cite{Nagele1997a,Rex2009}. They are likewise useful as input in the context of the empirically observed time-wavenumber factorization scaling of $S(q,t)$ at $q\approx q_m$ for intermediate to long correlation times \cite{Segre1996,Holmqvist2010}.

The intra-particle hydrodynamic structure has been accounted for using the hydrodynamic radius model where the particles are described hydrodynamically as no-slip spheres, characterized by a hydrodynamic radius derived from a single-particle transport property for unchanged direct interactions. In spite of its simplicity, the HRM is universally applicable since corrections terms are usually quite small, i.e. of quadratic order in the reduced slip length. In many particulate systems, these corrections are negligible, and a unique hydrodynamic radius $a_{h,f}$ can be used independent of the considered transport property. As we have shown in comparison with existing computer simulation results, the HRM based scaling expressions for hard-sphere-like particles are decent approximations also for strongly structured particles characterized by values of  
$\gamma$ significantly smaller than one, provided $a_h$ is deduced from an associated single-particle transport coefficient, namely from $[\eta]$ in viscosity calculations, and $D_0^t$ in self-diffusion and sedimentation  calculations for concentrated suspensions. While our focus has been on monodisperse particle systems, 
it is feasible to generalize the HRM to size polydisperse particles using an appropriate particle size distribution. 
In future work, it will be rewarding to search for possible extensions of the scaling relations 
to polydisperse colloidal systems and mixtures.

Most results presented in this work are for hydrodynamically structured, stiff particles with pure hard-core interactions where  the HRM reduces to the spherical annulus model. Part of our toolbox methods are likewise applicable, with appropriate modifications, to spherical particles with short-ranged soft interactions. We have illustrated this in Fig. \ref{fig:11_h_red_0_35_hertz_10_10000} for the reduced hydrodynamic function, $h_d(q)$, of soft Hertz potential particles with $\epsilon=10$. According to our calculations based on the $\delta\gamma$ method, hydrodynamic function scaling remains valid for soft spherical particles. Note here that the self-part corrected $\delta\gamma$ scheme, and the PA approximation for HRM particles, can be used for arbitrary pair potentials.

We finally point to the invalidity of the presented scaling relations for $D_S$, $K$, and $D_L$ 
for particles with long-range soft repulsion that is not adequately described 
by an effective excluded volume diameter. 
Examples in case are low salinity suspensions of charge-stabilized particles \cite{Heinen2010}, and quasi-two dimensional systems of magnetically repelling particles at a liquid-gas interface 
\cite{Zahn1997}. As shown both in computer simulations and experiments, $D_S$ and $K$ in these systems follow a fractal $\sim\phi^{4/3}$ and $\sim\phi^{1/3}$ concentration dependence, respectively, of zero and negative infinite initial slope \cite{Heinen2010,Banchio2008}. 
Moreover, the factorization scaling relation in Eqs. (\ref{eq:D_long_scaling}) 
and (\ref{eq:D_long_scaling_2}) predicting that 
$D_L < [D_L]_\text{no-HI}$ does not apply to these long-range repulsive systems where, on the contrary, 
a hydrodynamic enhancement of long-time self-diffusion is observed in theory \cite{Nagele1997a}, simulation \cite{Rinn1999,Nagele2002} 
and experiments \cite{Zahn1997}.  
The short-time self-diffusion coefficient in these systems is only weakly affected hydrodynamically 
at lower concentrations so that $D_S \approx D_0^t$. 
The major effect originates instead from the relaxation part $\Delta D$  
whose magnitude is lowered by the here dominating far-field part of the HIs \cite{Nagele1997a}.  
While the scaling expressions are invalid, the hydrodynamic structures of the low-salinity charge-
stabilized and magnetic particles is still well described by the HRM. On basis of the HRM the short-time properties of hydrodynamically structured charge-stabilized particles can be calculated using the self-part corrected $\delta\gamma$ scheme in conjunction with the PA method for the self-part.

\section*{Acknowledgements}
J.R. acknowledges support by the International Helmholtz Research School on Biophysics and Soft Matter
(IHRS BioSoft). G.N. and J.R. thank M. Heinen (CalTech, Pasadena), B. Cichocki (Warsaw University), and M. Ekiel-Jezewska and E. Wajnryb (Polish Academy of Sciences, Warsaw) for many helpful discussions. This work was partially funded by the Deutsche Forschungsgemeinschaft (DFG) within the framework of the SFB-985 (project B6).

\appendix
\section{Pairwise-additivity (PA) approximation applied to the HRM}
\label{app:pairwise_additive_approx}

The PA approximation of short-time diffusion properties is based on the cluster expansion of the $N$-particle translational mobility tensors of colloidal spheres,
\begin{align}
\bm{\mu}_{ij} \left({\bf X}\right) &= \mu_0^t \,\mathbb{1} \delta_{ij} + \left[ \bm{\mu}_{ij}^{(2)} \left({\bf X}\right) - \mu_0^t \, \mathbb{1}\delta_{ij}\right] \\ &+ \text{three-body terms} + \ldots\,,\notag 
\end{align}
where the three-body and higher-order cluster contributions are disregarded. Here, 
$\mu_0^t= D_0^\text{t}/(k_B T)$ is the single-particle translational mobility coefficient 
depending on the hydrodynamic particle structure, and $\mathbb{1}$ is the unit tensor.   
The hydrodynamic mobility tensors on the pairwise additive level,
\begin{align} \label{eq:pa_diff_tensor}
\bm{\mu}_{ij}^{(2)} \left({\bf X}\right) = \mu_0^t &\left[\delta_{ij} \left( \mathbb{1}+\sum_{l=1;l\neq i}^{N} \; \bm{\omega}_{11}\left(\mathbf{R}_{il}\right) \right)\right. \\ 
&\qquad\quad + \left(1-\delta_{ij}\right) \;\bm{\omega}_{12}\left(\mathbf{R}_{ij}\right) \Bigg] \,, \notag
\end{align}
with $\mathbf{R}_{ij}=\mathbf{R}_{i}-\mathbf{R}_{j}$, are fully accounted for including the near-contact lubrication terms. The two-particle tensors $\bm{\omega}_{11}$ and $\bm{\omega}_{12}$ describe the hydrodynamic self-interaction of a sphere through flow reflections at the second one, and cross-interactions of the two particles, respectively. The axial symmetry of the two-sphere problem allows for splitting these tensors into longitudinal and transversal components,
\begin{equation}
\bm{\omega}_{ij} = x_{ij}(r) \mathbf{\hat{r}}\mathbf{\hat{r}} + y_{ij} \left(r\right)\left[\mathbb{1}-\mathbf{\hat{r}}\mathbf{\hat{r}}\right] \,.
\label{eq:pa_tensors}
\end{equation}
The transversal and longitudinal mobilities $x_{ij}(r)$ and $y_{ij}(r)$, with $i,j \in \{1,2\}$, can be calculated recursively in the form of a power series in the reduced inverse pair distance $a_h/r$, 
combined with known near-contact lubrication expressions \cite{Jeffrey1984,Jones1988a,Kim1991}. 

Insertion of Eq. (\ref{eq:pa_diff_tensor}) into Eq. (\ref{eq:hyd_func_2}) leads to the PA approximation 
expressions for $D_S$ and $H_d(q)$ \cite{Heinen2011}. We present these here in a form suitable for HRM particles where one distinguishes between the hydrodynamic particle diameter, $\sigma_h=2 a_h$, 
and the direct interaction diameter $\sigma$. 
For rigid spherical particles, $\sigma=2a$ is the hard-core diameter, while for 
mechanically soft particles, $\sigma$ is identified with a characteristic soft diameter 
$\sigma_\text{s}$ such as the one in the Hertz potential.

Introducing the reduced length $x=r/\sigma_\text{h}$ and wavenumber $y=q\sigma_\text{h}$, 
the self-part contribution to $H(q)$ is 
\begin{align}
\left.{\frac{D_\text{S}\left(\phi,\gamma\right)}{D_0^t\left(\gamma\right)}}\right|_\text{PA} = 
1 + 8 \gamma^3 \phi \int_{0}^\infty&\;\!\text{d}x\;\! x^2\;\! g\left(\gamma x\right) \label{eq:pa_self}\\
&\cdot\left[ x_{11} \left(x\right) + 2 y_{11} \left(x\right) - 3 \right] \,. \notag
\end{align}
The distinct part contribution reads
\begin{align}\label{eq:pa_hyd_func_calc}
&\left.H^d\left(y,\gamma\right)\right|_\text{PA} = \gamma^3\phi   \\
&+18\gamma\phi\int_{0}^\infty\text{d}x\;xh\left(\gamma x\right)\left[j_0\left(xy\right)-\frac{j_1\left(xy\right)}{xy}
+  \gamma^2\frac{j_2\left(xy\right)}{6x^2}\right] \notag \\
&+ 24\gamma^3\phi \int_{0}^\infty\text{d}x\; x^2g\left(\gamma x\right) \overline{y}_{12}\left(x\right) 
j_0\left(xy\right)  \notag\\
&+ 24\gamma^3\phi \int_{0}^\infty \text{d}x \; x^2g\left(\gamma x\right) \left[ \overline{x}_{12}\left(x\right)-
\overline{y}_{12}\left(x\right) \right] \notag  \\
&\qquad\qquad\qquad\cdot\left[ \frac{j_1\left(xy\right)}{xy} - j_2\left(xy\right) \right] \,, \notag 
\end{align}
where $j_n$ is the spherical Bessel function of order $n$, and $h=g-1$ is the total correlation function. 
The overlines in $\overline{x}_{12}(x)$ and $\overline{x}_{12}(x)$ indicate that the respective far-field parts up to third order in $1/x$ have been subtracted off. The argument $\gamma x$ in $g$ and $h$, with $\gamma=\sigma_h/\sigma$, is a reminder that the RDF is commonly calculated as a function of $r/\sigma$ such as in the VW-PY solution for hard spheres.  

According to Eq. (\ref{eq:pa_self}), the first-order virial coefficient of $D_S$ is given by
\begin{align}
\lambda_\text{t}\left(\gamma\right) = 8\gamma^3\int_{0}^{\infty}& \text{d}x\; x^2 \exp\left[-\beta V\left(x\right)\right]\label{eq:sa_virial_exact} \\ 
&\cdot\left[x_{11}\left(x\right) +2y_{11}\left(x\right) -3\right] \,.\notag
\end{align}

The PA approximation expression for the high-frequency limiting viscosity, generalized to the HRM reads, 
\begin{align}
\left. \frac{\eta_\infty\left(\phi,\gamma\right)}{\eta_0}\right|_\text{PA} = 1&+\frac{5}{2}\gamma^3\phi\left[1+\gamma^3\phi\right] \\
 &+ 60\left(\gamma^3\phi\right)^2\int_{0}^\infty\text{d}x\, x^2 g\left(\gamma x\right)J\left(x\right) \,,\notag 
\end{align}
with the rapidly decaying two-sphere shear mobility function, $J(x)$, accounting for the two-body HIs. It decays asymptotically as $J(x) \sim(15/128)x^{-6}$. We employ an accurate numerical table for $J(x)$ based on recursion expressions and the lubrication analysis given in \cite{Jeffrey1992}. 

Finally, we note that in the spherical annulus model of hydrodynamically structured hard spheres, 
the lower integration boundary of all integrals in the present appendix is equal to $1/\gamma$.\\

\section{Self-part corrected Beenakker-Mazur method applied to the HRM}
\label{app:self-part-corrected-Beenakker-Mazur-method}

Except for $\eta_\infty$, the applicability range of the PA approximation is restricted to lower $\phi$ values where 
it becomes accurate. In contrast, the $\delta\gamma$-scheme for $H(q)$ and $\eta_\infty$ by Beenakker and Mazur 
\cite{Beenakker1983,Beenakker1984b} is applicable also to concentrated systems. We discuss here a standard version 
of this scheme, generalized to the HRM, where $S(q)$ is the only required input. 
Once the self-part $D_S$ of $H(q)$ is suitably corrected \cite{Westermeier2012,Heinen2011}, the 
$\delta\gamma$ method provides a decent description of short-time transport properties, 
for neutral and charge-stabilized particle systems alike (see \cite{Heinen2011}). 
The $\delta\gamma$ method results for $\eta_\infty$ and $H(q)$ reveal inaccuracies at all concentrations which can be partially attributed to its approximate treatment of the HIs. This is underlined in recent work by Makuch and Cichocki \cite{Makuch2012} where the approximation steps in the derivation of the $\delta\gamma$ scheme have been reduced. The fact that their revised version of the $\delta\gamma$ scheme with improved hydrodynamic mobility tensors does not significantly improve the agreement with simulation data for no-slip hard spheres points to a fortuitous cancellation of errors introduced in the approximate derivational steps of the original (non self-part corrected) Beenakker-Mazur method.

A simple yet significant improvement over the original $\delta\gamma$ scheme preserving its analytical simplicity 
is obtained from using this scheme for the distinct part $H_d(q)$ only, where it gives overall good results. This was shown in \cite{Banchio2008,Heinen2011,Westermeier2012} both for neutral and charge-stabilized particle systems.  
Regarding the self-part, $D_S$, of $H(q)$, accurate expressions can be used instead such as the scaling relation 
in Eq. (\ref{eq:self_diff_improved}) for particles with 
hard-core interactions, or the PA approximation expression in Eq. (\ref{eq:pa_self}) for lower-concentrated charge-stabilized systems. This hybrid procedure is referred to as the self-part corrected $\delta\gamma$ scheme.

The $\delta\gamma$ scheme expression for $H_d(q)$ by Beenakker and Mazur \cite{Beenakker1983,Beenakker1984b}, 
generalized to the HRM, reads
\begin{align}
&\left.H^d \left(y,\gamma\right)\right|_{\delta\gamma} = \label{eq:dg_annulus} \\
&\frac{3}{4\pi} \int_0^\infty \text{d}y' \; \left(\frac{\sin\left(\frac{1}{2}y'\right)}{\frac{1}{2}y'}\right)^2 \frac{1}{1+\gamma^3\phi S_{\gamma_0}\left(\frac{1}{2}y'\right)} \notag \\
\cdot&\int_{-1}^{1} \text{d}\mu\; \left(1-\mu^2\right) \left[S\left(\frac{1}{\gamma}\sqrt{y^2+y'^2-2yy'\mu}\right)-1\right]\,, \notag 
\end{align}
with $y=q\sigma_h$. 
The function $S_{\gamma_0}(x)$ consists of an infinite sum of wavenumber-dependent 
contributions depending on $\phi_h$ as well as on the inter-related scalar coefficients 
$\gamma_0^{(n)}$, with $n\in \{0,1,2,\cdots\}$. Explicit expressions for 
$S_{\gamma_0}(x)$ and $\gamma_0^{(n)}$ are given in 
\cite{Beenakker1984b,Genz1991}. We have calculated the $\gamma_0^{(n)}$ coefficients in an iterative procedure 
up to $n=10$, using a fine grid of volume fractions in $\left[0.01-0.5\right]$ of grid size $\Delta\phi=0.01$. 
This has resulted in an improved accuracy as compared to the original work by Beenakker and Mazur. However, the differences in $H_d(q)$ and $\eta_\infty$ are quite small, i.e. there is no more than 
a $3 \%$ difference.   

The $\delta\gamma$ scheme expression for $\eta_\infty$ by Beenakker \cite{Beenakker1984a}, 
adapted to the HRM, is given by
\begin{align}
\left.\frac{\eta_\infty\left(\phi,\gamma\right)}{\eta_0} \right|_{\delta\gamma} &= \frac{1}{\lambda^{(0)}\left(\phi,\gamma\right)+\lambda^{(2)}\left(\phi,\gamma\right)}\,, \\
\intertext{where}
\lambda^{(0)}\left(\phi,\gamma\right) &= \left[1+\frac{5}{2}\gamma^3\phi\tilde{\gamma}_0^{(2)}\right]^{-1} \\
\lambda^{(2)}\left(\phi,\gamma\right) &= \frac{15}{2\pi}\gamma^3\phi\left[\tilde{\gamma}^{(2)}_0\lambda^{(0)}\left(\phi,\gamma\right)\right]^2  \\
&\cdot \int_0^\infty \text{d}y \frac{j_1^2\left(\frac{1}{2}y\right)}{1+\gamma^3\phi S_{\gamma_0}\left(\frac{1}{2}y,\gamma^3\phi\right)}\left[S\left(\frac{y}{\gamma}\right)-1\right] \,. \notag
\end{align}
Here, $\tilde{\gamma}^{(2)}_0= \gamma^{(2)}_0/n$ with $n$ denoting the particle number density. The argument $y/\gamma$ in the static structure factor is used as a reminder 
that it is usually calculated as a function of $q\sigma$.

\bibliography{library}% Produces the bibliography via BibTeX.

\end{document}